\documentclass[conference,compsoc]{IEEEtran}
\usepackage{hyperref}       
\usepackage{url}            
\usepackage{amsfonts}       
\usepackage{graphicx}       
\usepackage{amsthm}
\usepackage{amsmath}
\usepackage{algpseudocode, algorithm}
\usepackage{subcaption}
\usepackage{diagbox}
\algnewcommand\INPUT{\item[\textbf{Input:}]}%
\algnewcommand\OUTPUT{\item[\textbf{Output:}]}%

\ifCLASSOPTIONcompsoc
  \usepackage[nocompress]{cite}
\else
  \usepackage{cite}
\fi
\ifCLASSINFOpdf
\else
\fi
\begin{document}
\newtheorem{theorem}{Theorem}[section]
\newtheorem{lemma}{Lemma}[section]
\newtheorem{definition}{Definiton}[section]
\newtheorem{corollary}{Corollary}[section]
\newtheorem{proposition}{Proposition}[section]
\newtheorem{conjecture}{Conjecture}[section]
\newtheorem{property}{Property}[section]
\newtheorem{example}{Example}

%
\title{Randomized Response with Gradual Release of Privacy Budget}

\author{\IEEEauthorblockN{Mingen Pan}
\IEEEauthorblockA{Independent Researcher \\
Email: mepan94@gmail.com}
}


%


\maketitle

\begin{abstract}

An algorithm is developed to gradually relax the Differential Privacy (DP) guarantee of a randomized response. The output from each relaxation maintains the same probability distribution as a standard randomized response with the equivalent DP guarantee, ensuring identical utility as the standard approach. The entire relaxation process is proven to have the same DP guarantee as the most recent relaxed guarantee.

The DP relaxation algorithm is adaptable to any Local Differential Privacy (LDP) mechanisms relying on randomized response. It has been seamlessly integrated into RAPPOR, an LDP crowdsourcing string-collecting tool, to optimize the utility of estimating the frequency of collected data. Additionally, it facilitates the relaxation of the DP guarantee for mean estimation based on randomized response. Finally, numerical experiments have been conducted to validate the utility and DP guarantee of the algorithm.

\end{abstract}


%
\IEEEpeerreviewmaketitle

\section{Introduction}

Artificial intelligence has developed rapidly in recent years, largely driven by the vast amounts of data collected everywhere. However, this has raised concerns about privacy, as personal data is often used to train AI models \cite{shokri2015privacy}. Differential privacy (DP), initially proposed by \cite{dwork2006differential}, is a technique that can be used to protect individual privacy during model training \cite{shokri2015privacy, abadi2016deep}. DP ensures that the addition or removal of a single entry in a dataset does not significantly change the output of a query. Later, a more stringent version of DP, known as Local Differential Privacy (LDP), was introduced by \cite{LDP}. LDP requires sufficient perturbation of individual data before aggregation. In contrast to the original version, LDP enables local data processing, eliminating the need for a trusted curator to aggregate data and preventing eavesdropping during data transmission \cite{hassan2019differential}. Theoretically, no third parties are able to acquire the original value of a user's data. As a result, tech giants, including Google \cite{rappor}, Apple \cite{apple_privacy}, and Microsoft \cite{ding2017collecting}, have embraced LDP as an integral part of their data processing protocols

The guarantee of an LDP query can be extended to an object, denoted as a privacy budget, which represents the largest DP guarantee of all the queries against the object. The privacy budget may change over time. For example, time-sensitive data like disease incidence may request stronger privacy protection when freshly generated, but the privacy requirement may relax as time progresses, resulting in an increase of privacy budget. Another scenario is a data market, where the privacy budget of a datum may be increased if the seller is willing to exchange their privacy for extra profit. Additional scenarios include privacy budget scheduling \cite{luo2021privacy}, which involves the gradual release of privacy budgets to ensure a fair allocation.

In all these scenarios, an analyst will ask a question - how do we optimally utilize the privacy budget if it is released gradually? This paper will answer this question under LDP. Since most advanced LDP algorithms.e.g., \cite{rappor, nguyen2016collecting, ye2019privkv} are built upon randomized response \cite{chaudhuri2020randomized}, the following article will purely focus on the randomized response. Suppose we have a group of objects, each with an initial privacy budget of $\epsilon_1$ and fully utilized by $\epsilon_1$-LDP randomized responses. As the privacy budgets are released to $\epsilon_2$, simply running $(\epsilon_2 - \epsilon_1)$-LDP randomized responses against the objects may result in a larger variance of the estimated frequency compared to an $\epsilon_2$-LDP randomized response. This issue has been discussed in prior literature (e.g., \cite{koufogiannis2015gradual, wang2019collecting, pan2023knowledge}), which motivates this paper to explore an algorithm to allow a randomized response to consume the gradually-released budget and generate outputs with same utility as standard randomized responses.

\subsection{Our Result}

This paper develops and proves an algorithm that can gradually relax a randomized response to larger DP guarantees, allowing for the consumption of more privacy budget. Each relaxation generates an output with the same probability distribution as a randomized response with the corresponding DP guarantee. Formally, if a randomized response is relaxed from $\epsilon_1$ to $\epsilon_2, ...$ to $\epsilon_n$, and $o_1, o_2, ..., o_n$ are the corresponding outputs, the algorithm satisfies the following properties:

\begin{property}
    For $\forall i \in [n]$, the probability distribution of $o_{i}$ is the same as that of an $\epsilon_{i}$-LDP randomized response, which is formulated as:
    \begin{equation}
        Pr(o_{i} = a | X = a) = \frac{e^{\epsilon_{i}}}{e^{\epsilon_{i}} + m - 1} \notag
    \end{equation}
    and
    \begin{equation}
    \begin{gathered}
        Pr(o_{i} = b| X = a) = \frac{1}{e^{\epsilon_{i}} + m - 1} \\
        \text{ s.t. } b \ne a \notag
    \end{gathered}
    \end{equation}
\end{property}

\begin{property}
    For $\forall i \in [n]$, the composition of $o_{\le i} = \{o_1, o_2, ..., o_{i} \}$ is $\epsilon_{i}$-LDP. That is, for any $i \in [n]$, any $a, b \in [m]^2$, and any $o_{\le i} \in [m]^{i}$,
    
    \begin{equation}
        Pr(o_{\le i} | X = a) \le e^{\epsilon_{i}} Pr(o_{\le i} | X = b) . \notag
    \end{equation}
\end{property}

The paper proceeds to explore the relationship between the DP guarantee of a relaxation and the privacy budget of an object. It discovers that while an individual relaxation may have a DP guarantee larger than the privacy budget, the DP guarantee of the entire relaxation process always remains within the budget. Furthermore, a comparison with the collusion-proof randomized response in \cite{xiao2009optimal} reveals that our DP relaxation algorithm is also collusion-proof.

In principle, the DP relaxation of randomized response is applicable to any mechanism based on randomized response. The first discussed application involves replacing the repeated noisy samplings in RAPPOR \cite{rappor}, an LDP crowdsourcing tool used for collecting strings. RAPPOR employs repeated noisy samplings to gradually reveal a hidden perturbation, and this process serves the same purpose as DP relaxation. The paper theoretically proves that substituting noisy samplings with DP relaxation enhances the utility of RAPPOR, bringing it to the optimal level as standard randomized responses. Additionally, the adoption of DP relaxation grants users full control over the relaxation of their privacy. The second application of DP relaxation pertains to the randomized-response-based mean estimation. Through the integration of DP relaxation, the DP guarantee of mean estimation can be relaxed while still making optimal use of the privacy budget. The final application discussed is serving as a fundamental component in a data market. In this scenario, users can sell their private data while adhering to their privacy budget. DP relaxation generates a sequence of randomized responses with varying DP guarantees, enabling users to sell and resell their data with different DP guarantees to different buyers, while preserving the overall privacy budget.

Finally, this paper conducts two experiments to validate the correctness and utility of the DP relaxation of randomized response. The first experiment concentrates on binary randomized responses, and the second one focuses on polychotomous randomized responses. Both experiments utilize the outputs of the relaxation to estimate the frequency of different values within the queried data. In both experiments, the results demonstrate that the accuracy and variance of the estimated frequency align with those obtained from standard randomized responses. Additionally, various inference methods are employed in both experiments to infer the true values of the queried data. Remarkably, all inference methods exhibit error rates not lower than the theoretical minimum guaranteed by LDP.

\subsection{Existing Work}

Prior research has predominantly delved into the relaxation of DP guarantees, often referred to as noise reduction, within the context of Central Differential Privacy. \cite{xiao2011ireduct} introduced the NoiseDown algorithm to relax the DP guarantee of the Laplace mechanism. However, \cite{xiao2022answering} identified an issue with the NoiseDown algorithm, suggesting that the algorithm in \cite{koufogiannis2015gradual} serves as the corrected version for relaxing the DP guarantee of the Laplace mechanism. This technique was adapted in empirical risk management (ERM) by \cite{ligett2017accuracy} to gradually determine the minimum DP guarantee satisfying the accuracy requirements of an ERM. Moreover, \cite{whitehouse2022brownian} developed an algorithm to gradually reduce the noise of the Gaussian mechanism, which was used to address accuracy-constrained ERM. The noise-reduction technique also extends to other applications, such as extracting the common ground of two different mechanisms \cite{xiao2022answering}.

As of now, there is no existing research on relaxing the guarantee of LDP queries. Notably, \cite{xiao2009optimal} created an algorithm to generate a sequence of randomized responses, ensuring that an adversary with all the responses can only learn as much as having the latest response. The increase in retention probability in this context is akin to relaxing the DP guarantee. However, this paper solely focused on collusion-proofing for Bayesian inference and did not establish any LDP guarantees for the entire process. A comparison between this algorithm and our work will be presented in Section \ref{sec:collusion_proof}.

Additionally, \cite{li2011enabling} developed an algorithm to reduce the Gaussian noise of individual data while preventing collusion. Similar to \cite{xiao2009optimal}, the algorithm focused on collusion-proofing and did not delve into LDP.

\section{Background}
\subsection{Local Differential Privacy (LDP)}

Local Differential Privacy (LDP) has been initially proposed in \cite{LDP} to ensure a query against an object does not significantly disclose its actual value. 

\begin{definition}
   A query $M$ is $\epsilon$-LDP, if and only if, for any $x \in \mathcal{X}, x' \in \mathcal{X}$, and any output $y$,

\begin{equation}
    Pr(M(x) = y | x) \le e^{\epsilon} Pr(M(x') = y | x') ,
\end{equation} 
\end{definition}

\noindent which is equivalent to, for any $y$ generated by $M$,

\begin{equation}
    \frac{\max\limits_{x \in \mathcal{X}} Pr(M(x) = y | x)}{\min\limits_{x' \in \mathcal{X}} Pr(M(x') = y | x')} \le e^{\epsilon}  .
    \label{eq:max_min_ldp_def}
\end{equation}

\noindent While there are other variations of LDP, this paper focuses only on the most widely-used form mentioned above.

\subsection{Privacy Budget}

A privacy budget limits the largest DP guarantee of all the queries against an object. It was originally defined for Central Differential Privacy \cite{rogers2016privacy, lecuyer2019privacy}, and \cite{pan2023knowledge} adapted it to LDP as follows:

\begin{definition}
    Object $X$ has a privacy budget $\epsilon$ if and only if, at any time, the executed results $o_{\le n} = \{o_1, o_2, ..., o_n \}$ of all the queries against $X$ are required to satisfy
    \begin{equation}
         \frac{\max\limits_x Pr(o_{\le n} | X = x)}{\min\limits_{x'} Pr(o_{\le n} | X = x')} \le e^{\epsilon} .
         \label{eq:privacy_budget}
    \end{equation}
\end{definition}

\noindent This is equivalent to considering the composition of all the queries against the object $X$ as an über query, and this über query is $\epsilon$-LDP (see Eq. \eqref{eq:max_min_ldp_def}). Here, the composition is fully-adaptive \cite{rogers2016privacy, pan2023knowledge}, where the choice of the next query depends on the result of the previous queries. That is

\begin{equation}
    Pr(o_{\le n} | X) = Pr(o_1 | X) \prod\limits_{2 \le i \le n} Pr(o_i | o_{\le i - 1}, X) .
\end{equation}

\subsection{Randomized Response}

Randomized Response Technique \cite{chaudhuri2020randomized} is a common approach to establish LDP \cite{dwork2014algorithmic}, and serves as a building block of many advanced LDP algorithms, e.g., \cite{rappor, nguyen2016collecting, ye2019privkv}. It is generally used to estimate the frequency of some properties inside a group of objects without the need to know the exact values of these objects. Let $\mathbf{X}$ be the group of interest with size $n$, and $X_i$ represents the $i$-th member of the group. $f$ is a deterministic function mapping from an object to the property of interest. The range (all possible values) of $f$ is denoted as $\mathcal{Y}$, and the possible values are finite and indexed arbitrarily as $y_1, y_2, ..., y_m$, where $m = |\mathcal{Y}|$.

Randomized response allows an object to randomly choose a value from $\mathcal{Y}$ and output it to preserve privacy. The probability of choosing a value is required to be

\begin{equation}
    Pr(y \in \mathcal{Y} | X) = \begin{cases}
        \frac{e^{\epsilon}}{e^{\epsilon + m - 1}} & \text{if } y = f(X) \\
        \frac{1}{e^{\epsilon + m - 1}} & \text{if } y \ne f(X) .
    \end{cases}
\end{equation}

\noindent The above probability leads to $\epsilon$-LDP \cite{wang2016using}.

Denote $Y_i$ as the output of a randomized response given $f(X_i)$, and $\mathbf{Y} = \{Y_1, Y_2, ..., Y_n\}$ is the collection of all the responses from the group, which will be used to estimate the original frequency of $f(X) = y$, where $y$ is an arbitrary value. If $f$ only outputs a binary value, the unbiased estimation of the frequency of $f(X) = b$, i.e., $\frac{1}{n} \sum_{1 \le i \le n} (f(X_i) = b)$, can be derived as \cite{chaudhuri2020randomized}:

\begin{equation}
    \pi_b = \frac{(e^\epsilon + 1) \lambda_b - 1}{e^\epsilon - 1} ,
    \label{eq:rr_frequency_estimate_binary}
\end{equation}

\noindent where $\pi_b$ is the corresponding estimator, and $\lambda_b$ is the frequency of $Y = b$, i.e., $\frac{1}{n} \sum_{1 \le i \le n} (Y_i = b)$.

If $f$ outputs more than two values, the frequency of $f(X) = y$, i.e., $\frac{1}{n} \sum_{1 \le i \le n} (f(X_i) = y)$, can be unbiasedly estimated as \cite{chaudhuri2020randomized}:

\begin{equation}
    \hat{H}_{f(\mathbf{X})} = P^{-1} H_{\mathbf{Y}},
    \label{eq:rr_frequency_estimate_poly}
\end{equation}

\noindent where $H_{\mathbf{Y}}$ is the histogram of $\mathbf{Y}$ with the $j$-th element representing the frequency of $Y = y_j$;  $\hat{H}_{f(\mathbf{X})}$ is the estimator of $H_{f(\mathbf{X})}$, which is the histogram of $f(\mathbf{X})$ with the $j$-th element representing the frequency of $f(X) = y_j$; $P^{-1}$ is the inverse matrix of $P$, which is a $m \times m$ matrix with the following values:

\begin{equation}
    P_{ij} = Pr(y_i | f(X) = y_j) .
\end{equation}

In the following section, we will drop the notion $f$ and assume $f(X) = X$ for convenience, which is equivalent to treating $f(X)$ as the true value of the object $X$. However, the theorems in this article are extendable to any $f$ as long as the range of $f$ is finite.

\section{DP Guarantee Relaxation of Randomized Response}
\label{sec:relaxation_proof}

In this section, we initially present a solution to relax the DP guarantee of a binary randomized response once, and then extend it to a polychotomous randomized response. Finally, we demonstrate that the solution for a single relaxation is also applicable to multiple-time relaxations.

\subsection{Binary Randomized Response with Single Relaxation}
\label{sec:binary_relax}

Let's consider an object $X$ with a binary value. Without loss of generality, we denote the value of $X$ as $a$, and its opposite value as $b$. The initial privacy budget of $X$ is denoted as $\epsilon_1$. An $\epsilon_1$-LDP randomized response has been applied to the object, producing an output $o_1$. Subsequently, the privacy budget is increased to $\epsilon_2$. Our goal is to relax the DP guarantee of the randomized response to $\epsilon_2$ by applying a new query to the object, resulting in output $o_2$. The relaxation should adhere to the following properties:

\begin{property}
    The probability distribution of $o_2$ is the same as that of an $\epsilon_2$-LDP randomized response, which is formulated as:
    \begin{equation}
        Pr(o_2 = a | X = a) = \frac{e^{\epsilon_2}}{e^{\epsilon_2} + 1}
        \label{eq:end_state_aa_binary}
    \end{equation}
    and
        \begin{equation}
        Pr(o_2 = b | X = a) = \frac{1}{e^{\epsilon_2} + 1} .
        \label{eq:end_state_ba_binary}
    \end{equation}
    \label{pro:end_state_binary}
\end{property}

\begin{property}
    The composition of the two queries is $\epsilon_2$-LDP. That is, for any $a, b \in \{0, 1\}^2$, and any $o_1, o_2 \in \{0, 1\}^2$,
    
    \begin{equation}
        Pr(o_1 o_2 | X = a) \le e^{\epsilon_2} Pr(o_1 o_2 | X = b) .
        \label{eq:ldp_composition_binary}
    \end{equation}
    \label{pro:ldp_composition_binary}
\end{property}

Now, let's explore the conditions leading to the above properties. Since $o_1$ is generated by an $\epsilon_1$-LDP randomized response, its probability distribution is known. The unknown variables in this context are the conditional probability of $o_2$ given $o_1$, denoted as $Pr(o_2 | o_1, X)$. For symmetry and interchangeability of $a$ and $b$, we require

\begin{equation}
    Pr(o_2 = a | o_1 = a, X = a) = Pr(o_2 = b | o_1 = b, X = b) ,
    \label{eq:symmetry_aaa==bbb}
\end{equation}

\noindent and

\begin{equation}
    Pr(o_2 = b | o_1 = b, X = a) = Pr(o_2 = a | o_1 = a, X = b) .
\end{equation}

\noindent Also, we have
\begin{multline}
    Pr(o_2 = b | o_1 = a, X = a) \\ = 1 - Pr(o_2 = a  | o_1 = a, X = a)
\end{multline}

\noindent and
\begin{multline}
    Pr(o_2 = a | o_1 = b, X = a) \\ = 1 - Pr(o_2 = b | o_1 = b, X = a) .
\end{multline}

\noindent Thus, the only unknown variables are $Pr(o_2 = a | o_1 = a, X = a)$ and $Pr(o_2 = b | o_1 = b, X = a)$, denoted as $p_a$ and $p_b$ respectively. Eq. \eqref{eq:end_state_aa_binary} in Property \ref{pro:end_state_binary} can be rewritten as
\begin{multline}
    Pr(o_2 = a | X = a) \\ = Pr(o_1 = a | X = a) p_a + Pr(o_1 = b | X = a) (1 - p_b) 
    \label{eq:end_state_aa_binary_rewrite}
\end{multline}

\noindent Eq. \eqref{eq:end_state_ba_binary} will be derived automatically when Eq. \eqref{eq:end_state_aa_binary} is satisfied because $Pr(o_2 = a | X = a) + Pr(o_2 = b | X = a) = 1$. 

Regarding the LDP requirement of composition (Property \ref{pro:ldp_composition_binary}), we observe that if the following equation
\begin{multline}
    Pr(o_1 = a, o_2 = a | X = a) \\ = e^{\epsilon_2} Pr(o_1 = a, o_2 = a | X = b) 
    \label{eq:ldp_composition_binary_sufficient_step_1}
\end{multline}

\noindent is satisfied, Property \ref{pro:ldp_composition_binary} will be satisfied for all $o_1$ and $o_2$ (see Appendix \ref{sec:proof_ldp_binary} for the proof). Consider the symmetry from Eq. \eqref{eq:symmetry_aaa==bbb}, we have 

\begin{equation}
    \text{Eq. \eqref{eq:ldp_composition_binary_sufficient_step_1}} =  e^{\epsilon_2} Pr(o_1 = b, o_2 = b | X = a) ,
\end{equation}

\noindent which can be rewritten in form of conditional probability:

\begin{equation}
    Pr(o_1 = a | X = a) p_a = e^{\epsilon_2} Pr(o_1 = b | X = a) p_b .
    \label{eq:ldp_composition_binary_sufficient}
\end{equation}

\noindent Substitute $Pr(o_1 = a | X = a) =  \frac{e^{\epsilon_1}}{e^{\epsilon_1} + 1}$ and $Pr(o_1 = b | X = a) = \frac{1}{e^{\epsilon_1} + 1}$ into Eq. \eqref{eq:end_state_aa_binary_rewrite} and \eqref{eq:ldp_composition_binary_sufficient}, and we have
 
\begin{equation}
    \frac{e^{\epsilon_1}}{e^{\epsilon_1} + 1} p_a + \frac{1}{e^{\epsilon_1} + 1} (1 - p_b) = \frac{e^{\epsilon_2}}{e^{\epsilon_2} + 1}
\end{equation}

\noindent and

\begin{equation}
    \frac{e^{\epsilon_1}}{e^{\epsilon_1} + 1} p_a = e^{\epsilon_2} \frac{1}{e^{\epsilon_1} + 1} p_b .
\end{equation}

\noindent Solving the above equation set, we have:

\begin{theorem}
    If 
    \begin{equation}
        p_a = Pr(o_2 = a | o_1 = a, X = a) = \frac{e^{\epsilon_2} - e^{-\epsilon_1}}{e^{\epsilon_2} - e^{-\epsilon_2}} 
        \label{eq:binary_pa}
    \end{equation}
    
    \noindent and
    
    \begin{equation}
        p_b = Pr(o_2 = b | o_1 = b, X = a) = \frac{e^{\epsilon_1 + \epsilon_2} - 1}{e^{2\epsilon_2} - 1} ,
        \label{eq:binary_pb}
    \end{equation}

     \noindent both Properties \ref{pro:end_state_binary} and \ref{pro:ldp_composition_binary} will be satisfied.

\end{theorem}

\subsection{Polychotomous Randomized Response with Single Relaxation}
\label{sec:single_relax_poly}

Now, we will investigate the scenario where an object receiving a randomized response has more than two possible values. Consider an object $X$ with $m$ possible values. For convenience, the $m$ possible values are assumed to be $\{1, 2, ..., m\}$, denoted as $[m]$, and the true value of $X$ is represented as $a \in [m]$. The initial privacy budget of $X$ is $\epsilon_1$, and an $\epsilon_1$-LDP randomized response has been applied, resulting in output $o_1$. Now, the privacy budget is relaxed to $\epsilon_2$, and our objective is to relax the DP guarantee of the randomized response to $\epsilon_2$ by applying a query, which outputs $o_2$. Similar to the binary case in Section \ref{sec:binary_relax}, the relaxation should adhere to the following properties:

\begin{property}
    The probability distribution of $o_2$ is the same as that of an $\epsilon_2$-LDP randomized response, which is formulated as:
    \begin{equation}
        Pr(o_2 = a | X = a) = \frac{e^{\epsilon_2}}{e^{\epsilon_2} + m - 1}
        \label{eq:end_state_aa_poly}
    \end{equation}
    and
    \begin{equation}
    \begin{gathered}
        Pr(o_2 = b| X = a) = \frac{1}{e^{\epsilon_2} + m - 1} \\
        \text{ s.t. } b \ne a
    \end{gathered}
        \label{eq:end_state_ba_poly}
    \end{equation}
    \label{pro:end_state_poly}
\end{property}

\begin{property}
    The composition of the two queries is $\epsilon_2$-LDP. That is, for any $a, b \in [m]^2$, and any $o_1, o_2 \in [m]^2$,
    
    \begin{equation}
        Pr(o_1 o_2 | X = a) \le e^{\epsilon_2} Pr(o_1 o_2 | X = b) .
    \end{equation}
    \label{pro:ldp_composition_poly}
\end{property}

Now, let's explore the number of variables when finding the conditions to satisfy the above properties. We will utilize symmetry as much as possible. If $o_1 = a$ and $X = a$, then $Pr(o_2 | o_1 = a, X = a)$ should be identical for every $o_2 \ne a$. Thus, if we denote

\begin{equation}
    Pr(o_2 = a | o_1 = a, X = a) = p_{aa} ,
\end{equation}

\noindent where the first letter of the subscript represents the value of $o_1$ and the second one represents $o_2$ (same for other variables below), then we have 
\begin{equation}
    \begin{gathered}
    Pr(o_2 = b | o_1 = a, X = a) = \frac{1 - p_{aa}}{m - 1} \\
    \text{ s.t. } b \ne a.
    \end{gathered}
\end{equation}

If $o_1 \ne a$, denoted as $o_1 = b$, then we should allow different conditional probability for $o_2$ to output $a$ (the true value) and $b$ (the value of $o_1$), but it should have identical probability to output other values different than $a$ and $b$, denoted as $c$. Thus, if we denote

\begin{equation}
    Pr(o_2 = a | o_1 = b, X = a) = p_{ba}
\end{equation}

\noindent and

\begin{equation}
    Pr(o_2 = b | o_1 = b, X = a) = p_{bb} ,
    \label{eq:def_bb_given_a}
\end{equation}

\noindent then we also have

\begin{equation}
    Pr(o_2 = c | o_1 = b, X = a) = \frac{1 - p_{ba} - p_{bb}}{m - 2} .
    \label{eq:def_bc_given_a}
\end{equation}

With the variables of $p_{aa}$, $p_{ba}$, and $p_{bb}$, we can rewrite Eq. \eqref{eq:end_state_aa_poly} as
\begin{multline}
    \sum\limits_{b \ne a} Pr(o_1 = b | X = a) p_{ba} +  \\ Pr(o_1 = a | X = a) p_{aa} = \frac{e^{\epsilon_2}}{e^{\epsilon_2} + m - 1} ,
\end{multline}

\noindent which is equivalent to

\begin{equation}
    \frac{m - 1}{e^{\epsilon_1} + m - 1} p_{ba} +  \frac{e^{\epsilon_1}}{e^{\epsilon_1} + m - 1} p_{aa} = \frac{e^{\epsilon_2}}{e^{\epsilon_2} + m - 1} .
    \label{eq:end_state_e2}
\end{equation}

\noindent Eq. \eqref{eq:end_state_ba_poly} will be derived automatically if Eq. \eqref{eq:end_state_aa_poly} is satisfied because of the identical treatment of the values other than the true value $a$ and $\sum\limits_{b \ne a} Pr(o_2 = b | X = a) +  Pr(o_2 = a | X = a) = 1$.

Now, let's express the LDP requirement for composition (Property \ref{pro:ldp_composition_poly}) in terms of $p_{aa}$, $p_{ba}$, and $p_{bb}$. Consider the scenario where $o_1 = o_2 = b$:
\begin{multline}
    Pr(o_1 = a, o_2 = a | X = a) \\ \le e^{\epsilon_2} Pr(o_1 = a, o_2 = a | X = b) .
\end{multline}

\noindent Given symmetry, we have $Pr(o_1 = a, o_2 = a | X = b) = Pr(o_1 = b, o_2 = b | X = a)$, leading to the equivalent form:
\begin{multline}
    Pr(o_1 = a, o_2 = a | X = a) \\ \le e^{\epsilon_2} Pr(o_1 = b, o_2 = b | X = a) .
\end{multline}

\noindent This further simplifies to:

\begin{equation}
    Pr(o_1 = a | X = a) p_{aa} \le e^{\epsilon_2} Pr(o_1 = b | X = a) p_{bb}.
\end{equation}

\noindent Given that $o_1$ is generated by an $\epsilon_1$-LDP randomized response, we have $Pr(o_1 = a | X = a) = \frac{e^{\epsilon_1}}{e^{\epsilon_1} + m - 1}$ and $Pr(o_1 = b | X = a) = \frac{1}{e^{\epsilon_1} + m - 1}$ where $b \ne a$. Substituting these into the above equation, we get:

\begin{equation}
    e^{\epsilon_1} p_{aa} \le e^{\epsilon_2} p_{bb}.
    \label{eq:aa_bb}
\end{equation}

The symmetry of LDP definition also requires:
\begin{multline}
    Pr(o_1 = a, o_2 = a | X = b) \\ \le e^{\epsilon_2} Pr(o_1 = a, o_2 = a | X = a) ,
\end{multline}

\noindent which is equivalent to
\begin{multline}
    Pr(o_1 = b, o_2 = b | X = a) \\ \le e^{\epsilon_2} Pr(o_1 = a, o_2 = a | X = a) ,
\end{multline}

\noindent and can be simplified as

\begin{equation}
    p_{bb} \le e^{\epsilon_2}  e^{\epsilon_1} p_{aa}
    \label{eq:bb_aa}
\end{equation}

\noindent in a similar way. Similar derivations are done for other scenarios of $o_1$ and $o_2$:

\begin{center}
\begin{tabular}{|c | c |} 
 \hline
 $o_1$ & $o_2$ \\ [0.5ex] 
 \hline
 a & b   \\  
\hline
 b & a   \\  
\hline
 a & c   \\  
\hline
 b & c   \\  
\hline
 c & a   \\  
\hline
 c & b   \\  
\hline
\end{tabular}
\end{center}

\noindent where $c \in [m]$ is defined to be $c \ne a$ and $c \ne b$. The above scenarios lead to the following inequalities:

\begin{equation}
    e^{\epsilon_1} \frac{1 - p_{aa}}{m - 1} \le e^{\epsilon_2} p_{ba} ,
    \label{eq:ab_ba}
\end{equation}

\begin{equation}
    p_{ba} \le e^{\epsilon_2} e^{\epsilon_1} \frac{1 - p_{aa}}{m - 1} ,
    \label{eq:ba_ab}
\end{equation}

\begin{equation}
    e^{\epsilon_1} \frac{1 - p_{aa}}{m - 1} \le e^{\epsilon_2} \frac{1 - p_{ba} - p_{bb}}{m - 2},
    \label{eq:ac_bc}
\end{equation}

\begin{equation}
   \frac{1 - p_{ba} - p_{bb}}{m - 2}  \le e^{\epsilon_2} e^{\epsilon_1} \frac{1 - p_{aa}}{m - 1},
   \label{eq:bc_ac}
\end{equation}

\begin{equation}
   p_{ba} \le e^{\epsilon_2} \frac{1 - p_{ba} - p_{bb}}{m - 2},
   \label{eq:ca_cb}
\end{equation}

\begin{equation}
   \frac{1 - p_{ba} - p_{bb}}{m - 2}  \le e^{\epsilon_2} p_{ba},
   \label{eq:cb_ca}
\end{equation}

\noindent respectively. The detailed derivation is presented in Appendix \ref{sec:detailed_ldp_proof}. Notably, there is one more scenario where $o_1 = c$ and $o_2 = d$, where $d \ne a$ and $d \ne b$. However, it is easy to see 
\begin{multline}
    Pr(o_1 = c, o_2 = d | X = a) \\ \le  e^{\epsilon_2} Pr(o_1 = c, o_2 = d | X = b)
\end{multline}

\noindent is always satisfied due to the symmetry defined in Eq. \eqref{eq:def_bb_given_a} and \eqref{eq:def_bc_given_a}, which leads to
\begin{multline}
    Pr(o_2 = d | o_1 = c, X = a) \\ = Pr(o_2 = d | o_1 = c, X = b) .
\end{multline}

\noindent Thus, this scenario is ignored.

Additionally, $p_{aa}$, $p_{ba}$, and $p_{bb}$ should satisfy the basic characteristic of probability, which is formulated as

\begin{equation}
    \begin{gathered}
        0 \le p_{aa} \le 1 , \\
        0 \le p_{ba} \le 1, \\
        0 \le p_{bb} \le 1, \\
        \text{and } 0 \le p_{bb} + p_{ba} \le 1 .
    \end{gathered}
    \label{eq:basic_prob}
\end{equation}

Ultimately, our problem is to find the solution for $p_{aa}$, $p_{ba}$, and $p_{bb}$ that satisfies Eq. \eqref{eq:end_state_e2} and Inequalities \eqref{eq:aa_bb}, \eqref{eq:bb_aa}, \eqref{eq:ab_ba} - \eqref{eq:cb_ca}, and \eqref{eq:basic_prob}. We have determined that when

\begin{align}
    & p_{aa} =\frac{e^{\epsilon_2}}{e^{\epsilon_2} - 1} - \frac{ e^{\epsilon_2 - \epsilon_1} (e^{\epsilon_1}  + m - 1)}{(e^{\epsilon_2} - 1)(e^{\epsilon_2} + m - 1)} \notag \\
    & p_{ba} =  \frac{ e^{2 \epsilon_2} - e^{\epsilon_1 + \epsilon_2 }}{(e^{\epsilon_2} - 1)(e^{\epsilon_2} + m - 1)} \notag  \\
    & p_{bb} =\frac{e^{\epsilon_1}}{e^{\epsilon_2} - 1} - \frac{e^{\epsilon_1}  + m - 1}{(e^{\epsilon_2} - 1)(e^{\epsilon_2} + m - 1)} ,
    \label{eq:solution_poly}
\end{align}

\noindent the above (in)equalities are satisfied. The approach to finding this solution and its proof are presented in Appendix \ref{sec:feasibility}. Formally, we have the following theorem:

\begin{theorem}
    Assume that $o_1$ is generated from an $\epsilon_1$-LDP randomized response. If $Pr(o_2 | o_1, X)$ are set to be $p_{aa}$, $p_{ba}$, and $p_{bb}$, then $o_2$ will relax the DP guarantee of $o_1$ to $\epsilon_2$.
    \label{thm:single_relax_poly}
\end{theorem}

Tables \ref{tab:p_aa_example} - \ref{tab:p_ba_example} provide examples of $p_{aa}$, $p_{ba}$, and $p_{bb}$ with DP guarantees gradually relaxed from $\epsilon = 0.1$ to $\epsilon = 2$, and the domain size $m$ ranges from 3 to 10.

Additionally, setting $m = 2$ in Eq. \eqref{eq:solution_poly} shows its equivalence to Eq. \eqref{eq:binary_pa} and \eqref{eq:binary_pb}.

\subsection{Continual Release of Privacy Budget}

The previous section introduces the algorithm to relax the DP guarantee of a randomized response once, and we would like to extend it to relax the DP guarantee multiple times. Similar to Section \ref{sec:single_relax_poly}, we still consider the polychotomous randomized response against an object $X$ with $m$ possible values. The true value of $X$ is denoted as $a$. An $\epsilon_1$-LDP randomized response was applied to $X$, producing output $o_1$. The privacy budget of $X$ will then be gradually released $n - 1$ times, with privacy budgets after each release denoted as $\epsilon_2, \epsilon_3, ..., \epsilon_n$. The DP guarantee of the randomized response will be relaxed accordingly, and the corresponding outputs are denoted as $o_2, o_3, ..., o_n$. The entire relaxation process should have the following properties:

\begin{property}
    For $\forall i \in [n]$, the probability distribution of $o_{i}$ is the same as that of an $\epsilon_{i}$-LDP randomized response, which is formulated as:
    \begin{equation}
        Pr(o_{i} = a | X = a) = \frac{e^{\epsilon_{i}}}{e^{\epsilon_{i}} + m - 1}
        \label{eq:end_state_aa_continual}
    \end{equation}
    and
    \begin{equation}
    \begin{gathered}
        Pr(o_{i} = b| X = a) = \frac{1}{e^{\epsilon_{i}} + m - 1} \\
        \text{ s.t. } b \ne a
    \end{gathered}
        \label{eq:end_state_ba_continual}
    \end{equation}
    \label{pro:end_state_continual}
\end{property}

\begin{property}
    For $\forall i \in [n]$, the composition of $o_{\le i} = \{o_1, o_2, ..., o_{i} \}$ is $\epsilon_{i}$-LDP. That is, for any $i \in [n]$, any $a, b \in [m]^2$, and any $o_{\le i} \in [m]^{i}$,
    
    \begin{equation}
        Pr(o_{\le i} | X = a) \le e^{\epsilon_{i}} Pr(o_{\le i} | X = b) .
        \label{eq:ldp_o_<=i}
    \end{equation}
    \label{pro:ldp_composition_continual}
\end{property}

We will prove that $p_{aa}$, $p_{ba}$, and $p_{bb}$ in Section \ref{sec:single_relax_poly} still satisfy the above properties. Here, $p_{aa}$, $p_{ba}$, and $p_{bb}$ are updated to represent $Pr(o_{i} | o_{i-1}, X)$ instead of $Pr(o_2 | o_1, X)$; $\epsilon_1$ and $\epsilon_2$ are replaced by $\epsilon_{i-1}$ and $\epsilon_{i}$, respectively. That is,

\begin{align}
    & p_{aa} =\frac{e^{\epsilon_{i}}}{e^{\epsilon_{i}} - 1} - \frac{ e^{\epsilon_i - \epsilon_{i-1}} (e^{\epsilon_{i-1}}  + m - 1)}{(e^{\epsilon_i} - 1)(e^{\epsilon_i} + m - 1)} \notag \\
    & p_{ba} =  \frac{ e^{2 \epsilon_i} - e^{\epsilon_{i-1} + \epsilon_i }}{(e^{\epsilon_i} - 1)(e^{\epsilon_i} + m - 1)} \notag  \\
    & p_{bb} =\frac{e^{\epsilon_{i-1}}}{e^{\epsilon_i} - 1} - \frac{e^{\epsilon_{i-1}}  + m - 1}{(e^{\epsilon_i} - 1)(e^{\epsilon_i} + m - 1)} .
    \label{eq:solution_poly_i++}
\end{align}

\noindent Therefore, the updated version of Theorem \ref{thm:single_relax_poly} will be

\begin{theorem}
        Assume that $o_{i-1}$ is generated from an $\epsilon_{i-1}$-LDP randomized response. If $Pr(o_{i} | o_{i-1}, X)$ is set to be $p_{aa}$, $p_{ba}$, and $p_{bb}$ in Eq. \eqref{eq:solution_poly_i++}, then $o_{i-1}$ and $o_{i}$ will satisfy the following properties:
        \label{thm:relax_i++}
\end{theorem}

\begin{property}
    The probability distribution of $o_{i}$ is the same as that of an $\epsilon_{i}$-LDP randomized response, which has the same form as Eq. \eqref{eq:end_state_aa_continual} and \eqref{eq:end_state_ba_continual}.
    \label{pro:end_state_i++}
\end{property}

\begin{property}
    The composition of $o_{i-1}$ and $o_{i}$ is $\epsilon_{i}$-LDP. That is, for any $a, b \in [m]^2$, and any $o_{i-1}, o_{i} \in [m]^2$,
    \begin{equation}
        Pr(o_{i-1} o_{i} | X = a) \le e^{\epsilon_{i}} Pr(o_{i-1} o_{i}  | X = b) .
        \label{eq:ldp_composition_i++}
    \end{equation}
    \label{pro:ldp_composition_i++}
\end{property}

\noindent The proof of Theorem \ref{thm:relax_i++} is omitted because it will be duplicate as Section \ref{sec:single_relax_poly}. Property \ref{pro:end_state_continual} requires Property \ref{pro:end_state_i++} to be true for every $i \in [n]$. Notably, the assumption of Theorem \ref{thm:relax_i++} is equivalent to Property \ref{pro:end_state_i++} with $i$ replaced by $i - 1$. Therefore, we can use the mathematical induction to prove Property \ref{pro:end_state_continual}: Theorem \ref{thm:relax_i++} indicates that Property \ref{pro:end_state_i++} is true for case $i$ if it is true for case $i - 1$. Also, when $i = 1$, Property \ref{pro:end_state_i++} is true because $o_1$ is generated by an $\epsilon_1$-LDP randomized response. Therefore, Property \ref{pro:end_state_i++} is true for every $i \ge 1$, and we have proved the following theorem:

\begin{theorem}
    If $Pr(o_i | o_{i - 1}, X)$ of $o_2, ..., o_n$ are set to be $p_{aa}$, $p_{ba}$, and $p_{bb}$ in Eq. \eqref{eq:solution_poly_i++}, Property \ref{pro:end_state_continual} will be satisfied.
\end{theorem}

Next, we will prove the $p_{aa}$, $p_{ba}$, and $p_{bb}$ also satisfy Property \ref{pro:ldp_composition_continual}. Before it, we need to prove a lemma 

\begin{lemma}
For any $a, b \in [m]^2$, and any $o_{\le i} \in [m]^i$,
    \begin{equation}
        \frac{Pr(o_{\le i} | X = a)}{Pr(o_{\le i} | X = b)} = \frac{Pr(o_i | X = a)}{Pr(o_i | X = b)}
        \label{eq:xxo==o}
    \end{equation}
    is satisfied for $i \ge 1$.
    \label{lem:xxo==o}
\end{lemma}

\noindent Proof: we will use mathematical induction, and assume Eq. \eqref{eq:xxo==o} is correct when $i = k$ and $k \ge 1$. Then, we will expand the left hand side of Eq. \eqref{eq:xxo==o} when $i = k + 1$:
\begin{multline}
     \frac{Pr(o_{\le k + 1} | X = a)}{Pr(o_{\le k + 1} | X = b)} \\ =  \frac{Pr(o_{\le k} | X = a) Pr(o_{k+1} | o_{\le k}, X = a)}{Pr(o_{\le k} | X = a) Pr(o_{k+1} | o_{\le k}, X = b)}
     \\ =  \frac{Pr(o_{k} | X = a) Pr(o_{k+1} | o_{k}, X = a)}{Pr(o_{k} | X = a) Pr(o_{k+1} | o_{k}, X = b)} \\ = \frac{Pr(o_k, o_{k+1} | X = a)}{Pr(o_k, o_{k+1} | X = b)} .
\end{multline}

\noindent Now the question becomes proving

\begin{equation}
    \frac{Pr(o_k, o_{k+1} | X = a)}{Pr(o_k, o_{k+1} | X = b)} = \frac{Pr(o_{k+1} | X = a)}{Pr(o_{k+1} | X = b)} .
\end{equation}

\noindent Since there are 9 possibles case for $(o_k, o_{k+1})$, where each of them can be either $a$, $b$, and $c$ ($c \ne a$ and $c \ne b$). The equation can be proved by substituting $Pr(o_k, o_{k+1} | X)$ in the form of $\epsilon_{k}, \epsilon_{k+1}, m$ into all these cases. Use $o_1 = o_2 = a$ as example, where

\begin{equation}
    \frac{Pr(a, a | X = a)}{Pr(a, a | X = b)} = e^{\epsilon_2} = \frac{Pr(a | X = a)}{Pr(a | X = b)} .
\end{equation}

\noindent Other cases can be proved similarly. Furthermore, when $k = 1$, Eq. \eqref{eq:xxo==o} is automatically satisfied because $o_{\le 1} = o_1$. $\qed$

Now, Property \ref{pro:ldp_composition_continual} can be proved using Lemma \ref{lem:xxo==o} and the LDP of a standard randomized response:

\begin{equation}
    \frac{Pr(o_{\le i} | X = a)}{Pr(o_{\le i} | X = b)} = \frac{Pr(o_{i} | X = a)}{Pr(o_{i} | X = b)} \le e^{\epsilon_{i}} . \qed
\end{equation}

\noindent Thus, we have established that

\begin{theorem}
    If $Pr(o_i | o_{i - 1}, X)$ of $o_2, ..., o_n$ are set to be $p_{aa}$, $p_{ba}$, and $p_{bb}$ in Eq. \eqref{eq:solution_poly_i++}, Property \ref{pro:ldp_composition_continual} will be satisfied.
    \label{thm:proof_ldp_composition_continual}
\end{theorem}

With the above theoretical proof, the continual relaxation of the DP guarantee of a randomized response can be formulated as Algorithm \ref{algo:dp_relax}.

\begin{algorithm}
    \caption{Continual DP Relaxation of Randomized Response}
    \label{algo:dp_relax}
  \begin{algorithmic}[1]
    \INPUT Object $X$, initial DP guarantee $\epsilon_1$
    \State Apply $\epsilon_1$-LDP randomized response against $X$ and generate the result $o_1$.
    \State $i = 2$.
    \Loop
        \State Waiting for the next DP guarantee to relax, and denote the guarantee to be $\epsilon_i$. 
        \State Set $Pr(o_i | o_{i-1}, X)$ to be Eq. \eqref{eq:solution_poly_i++} .
        \State Generate $o_i$ given $Pr(o_i | o_{i-1}, X)$ and $o_{i-1}$.
        \State $o_i$ is sent to the server as the output of an $\epsilon_i$-LDP randomized response.
        \State $i++$.
    \EndLoop
  \end{algorithmic}
\end{algorithm}

\section{Discussion}
\subsection{DP Guarantee vs Privacy Budget}

Consider a DP relaxation of a binary randomized response from $\epsilon_1$ to $\epsilon_2$.  Eq. \eqref{eq:binary_pa} and \eqref{eq:binary_pb} have provided the solution for $Pr(o_2 | o_1, X)$. If we consider $o_2$ as a query against $X$, we can calculate its DP guarantee $\epsilon_\Delta$ accordingly:
\begin{multline}
    e^{\epsilon_\Delta} = \max \frac{ Pr(o_2 | o_1, X = a)}{Pr(o_2 | o_1, X = b)} =  \max 
    \begin{cases}
        \frac{p_a}{p_b} & \text{if } o_1 = o_2 , \\
        \frac{1 - p_a}{1 - p_b} & \text{if } o_1 \ne o_2 ,
    \end{cases}
    \\ = \max \{ e^{\epsilon_2 - \epsilon_1}, e^{\epsilon_2 + \epsilon_1} \} = e^{\epsilon_2 + \epsilon_1} 
\end{multline}

Suppose $\epsilon_1 = 1$ and $\epsilon_2 = 2$; $e^{\epsilon_\Delta}$ is calculated to be 3. If we calculate the DP guarantee of $o_1$ and $o_2$ using basic composition (i.e., simply add up their DP guarantees), the composed DP guarantee will be 4, which is obviously larger than 2. This phenomenon has been discussed in \cite{pan2023knowledge}, which has found that basic composition may overestimate the privacy loss of composed queries, and Eq. \eqref{eq:privacy_budget} should be the gold standard to determine if the executed queries against an object have already exhausted the privacy budget.

One may ask whether $e^{\epsilon_\Delta} = 3$ overuses the privacy budget of the queried object. The answer is no. The privacy budget of an object has no requirement for the DP guarantee of a single query but the DP guarantee of all the composed queries. Here is an intuitive explanation of why $e^{\epsilon_\Delta} = 3$ is allowed: this occurs only when $o_1 \ne o_2$. When $o_1$ is observed, an adversary thinks the object is \textit{somehow} likely to have the value of $o_1$. When $o_2$ is observed, the adversary is changed to believe that the object is \textit{more} likely to have the value of $o_2$. This degree of change from $o_1$ to $o_2$ is even larger than changing the belief from knowing nothing to $o_2$. However, the eventual confidence about the object is the same as directly observing an $\epsilon_2$-LDP query, so the overall DP guarantee is still $\epsilon_2$.

\subsection{Compare with Collusion-proof Perturbation}
\label{sec:collusion_proof}

\cite{xiao2009optimal} developed an algorithm to generate a sequence of randomized responses with gradually increased retention probability of the original value, while ensuring that an adversary with all responses cannot infer the original object better than knowing the last response, a property referred to as collusion-proof. The gradual increase in retention probability is similar to relaxing DP guarantees, and their work was derived from the collusion-proof requirement, whereas ours is derived from LDP. We have found that our algorithm can lead to the collusion-proof property, formally defined as:

\begin{theorem}
    For $\forall o_{\le n} \in [m]^n$ from the relaxation process, 
    \begin{equation}
        Q(X | o_{\le n} ) = Q(X | o_n) 
        \label{eq:collusion_proof}
    \end{equation}
    is satisfied.
    \label{thm:collusion_proof}
\end{theorem}

\noindent where $Q(X | ...)$ represents the Bayesian belief in the value of $X$ with the given information. The theorem can be derived from Lemma \ref{lem:xxo==o}, and the proof is presented in Appendix \ref{sec:proof_collusion_proof}. We conjecture that our algorithm could be equivalent to the collusion-proof perturbation \cite{xiao2009optimal}, assuming their algorithm was derived from the collusion-proof property. However, when substituting $\epsilon_1 = 1$ and $\epsilon_2 = 2$ (corresponding to $p_a = 0.36$ and $p_b = 0.68$ in their notation) into Eq. (7) of \cite{xiao2009optimal}, we realize their $Pr(o_2 | o_1, X)$ is different from ours, and the corresponding marginal probability of $Pr(o_2 | X)$ is different from that of an $\epsilon_2$-LDP randomized response. Since our result has been proved theoretically and numerically in Sections \ref{sec:relaxation_proof} and \ref{sec:experiment}, respectively, we recommend using Eq. \eqref{eq:solution_poly_i++} in this paper to relax the DP guarantee (i.e., retention probability) of randomized responses, which also provides collusion-proofness.

\section{Application}

\subsection{Gradual Privacy Relaxation in RAPPOR}

\begin{figure*}[ht!]
\centering
\subfloat[]{\label{fig:dp_rappor} \includegraphics[width=0.32\textwidth]{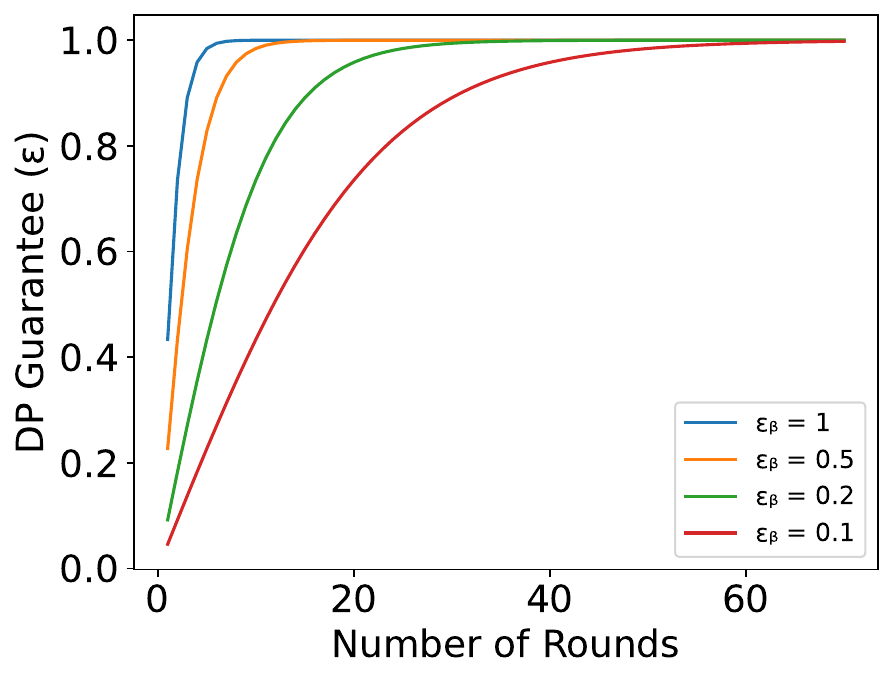}}
\hfill
\subfloat[]{\label{fig:var_rappor_full} \includegraphics[width=0.32\textwidth]{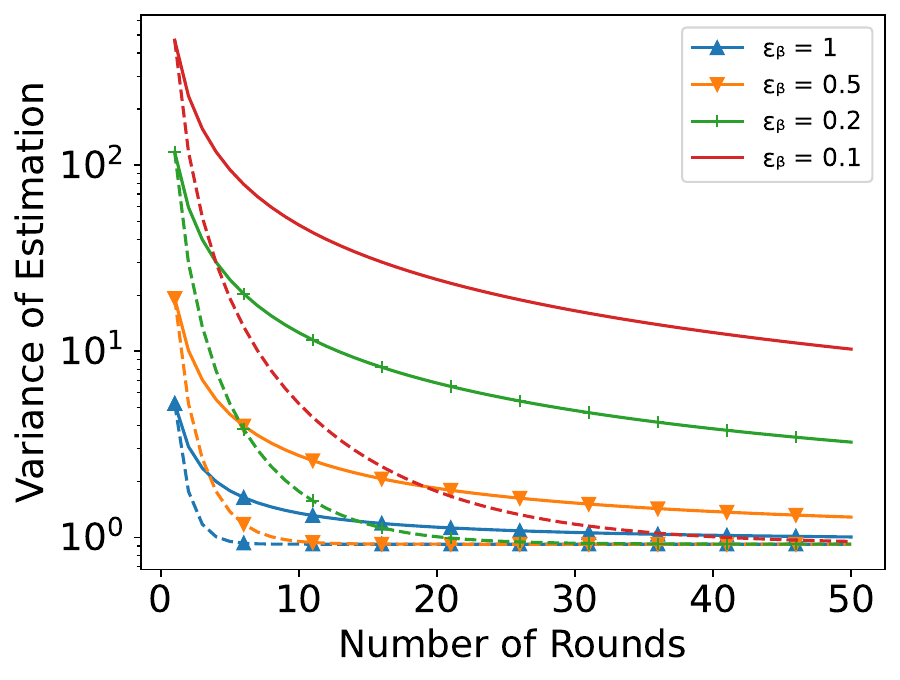}}%
\hfill
\subfloat[]{\label{fig:var_rappor_linear} \includegraphics[width=0.32\textwidth]{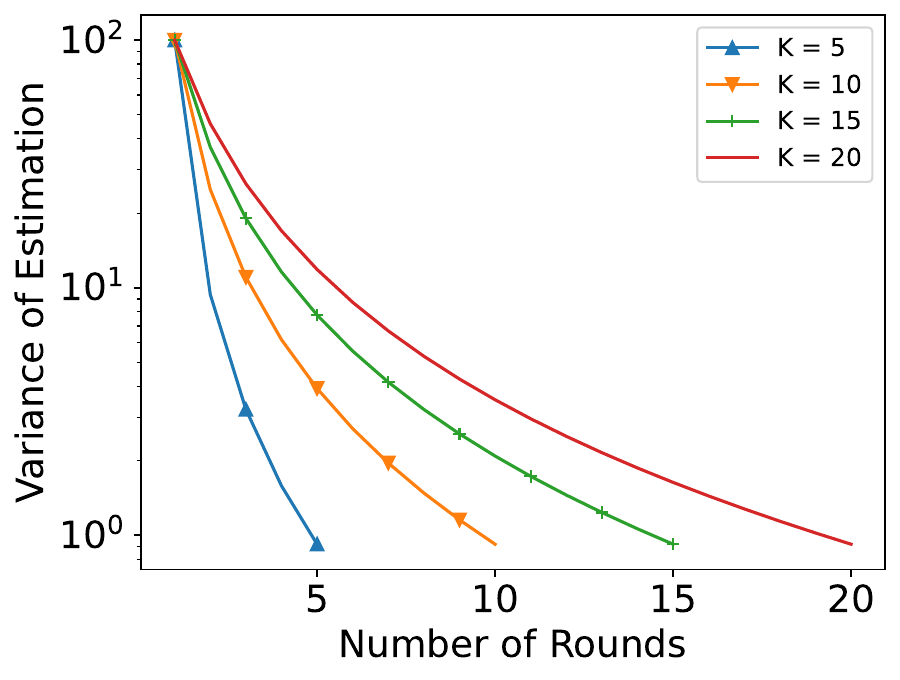}}%
\caption{(a) illustrates the DP guarantee of repeated noisy samplings with $\epsilon_\alpha = 1$ and varying $\epsilon_\beta$ as indicated in the legend; (b) depicts the variance in the estimated frequency of the original bit $B = 1$ with $\epsilon_\alpha = 1$ and various $\epsilon_\beta$ values specified in the legend. Solid curves represent repeated noisy samplings, while dashed curves depict the DP relaxation of randomized responses; (c) showcases the variance in the estimated frequency of the original bit $B = 1$ derived from randomized responses with DP guarantee $\epsilon$ relaxed from 0.1 to 1.0 linearly. Different curves denote different rounds to complete relaxation, labeled as $K$. Both (b) and (c) use $N = 1$ when calculating the variance.}
\end{figure*}

RAPPOR \cite{rappor} is an LDP crowdsourcing tool developed by Google to collect strings from clients and analyze the frequency of the strings. A client processes a string in three steps before sending it to the RAPPOR server: \newline

1. Hash a string into a Bloom filter $\mathbf{B}$. \newline

2. Permanent Randomized Response: for each bit $B$ in the Bloom filter $\mathbf{B}$, apply an randomized response and output $B'$ such that

\begin{equation}
    B' = 
    \begin{cases}
        B & \text{w.p. } \alpha  \\
        \neg B & \text{w.p. } 1 - \alpha  
    \end{cases} ,
\end{equation}

\noindent where w.p. represents "with probability of". This is equivalent to a $\ln \frac{\alpha}{1 - \alpha}$-LDP randomized response against each bit. The perturbed bloom filter $\mathbf{B}'$ is stored on the client.
\newline

3. Noisy Sampling (originally named as Instantaneous Randomized Response in \cite{rappor}): allocate a new string $\mathbf{S}$ with the same size as the perturbed Bloom filter $\mathbf{B}'$, and each bit $S$ of $\mathbf{S}$ is assigned as follows:

\begin{equation}
    Pr(S = 1) = 
    \begin{cases}
        p & \text{if } B' = 1 \\
        q & \text{if } B' = 0 \\
    \end{cases} .
\end{equation}
\newline

4. Continually run Step 3 and send the generated string $\mathbf{S}$ to the server.
\newline

The purpose of noisy sampling (Step 3) is to avoid the immediate release of the perturbed Bloom filter $\mathbf{B}'$, thereby further protecting users' privacy. As more and more strings from the noisy sampling are reported, the server gradually gains a clearer picture of the actual value of the perturbed Bloom filter $\mathbf{B}'$. Thus, Steps 3 and 4 serve the same function of gradually relaxing the DP guarantee of the permanent randomized response (Step 2). Therefore, we can integrate the gradual relaxation of the randomized response into RAPPOR. Instead of repeatedly running noisy samplings, the DP guarantee of the permanent randomized response can be gradually relaxed from a small value to $\ln \frac{\alpha}{1 - \alpha}$-LDP for each bit.

We would like the DP guarantee of the permanent randomized response to be relaxed at the same rate as that of repeated noisy samplings. Here, we will calculate the DP guarantee of the repeated noisy samplings. Since each bit is independent in RAPPOR, the following discussion will only focus on one bit. To simplify the comparison, we assume $\beta = p = 1 - q$ in noisy samplings. Thus, the noisy sampling becomes a standard $\ln \frac{\beta}{1 - \beta}$-LDP randomized response. Subsequently, we can prove the DP guarantee after reporting $K$ times of noisy samplings, denoted as $\epsilon_{ns}(K)$, equals

\begin{equation}
   \epsilon_{ns}(K) = \ln \frac{e^{\epsilon_\alpha} e^{K \epsilon_\beta} + 1}{e^{\epsilon_\alpha} + e^{K \epsilon_\beta}} .
   \label{eq:eps_noisy_sampling}
\end{equation}

\noindent where $\epsilon_\alpha = \ln \frac{\alpha}{1 - \alpha}$ and $\epsilon_\beta = \ln \frac{\beta}{1 - \beta}$. The proof is at Appendix \ref{sec:dp_noisy_sampling}. When $K \rightarrow \infty$, $\epsilon_{ns}(K) \rightarrow \epsilon_\alpha$. The plot of $\epsilon_{ns}(K)$ vs $K$ is presented in Fig. \ref{fig:dp_rappor}, where $\epsilon_\alpha = 1.0$ and $\epsilon_\beta$ ranges from 0.1 to 1.0. 

Now, let's compare the utility of repeated noisy samplings with our proposed DP relaxation. The decoding process of RAPPOR involves estimating the frequency of each bit generated by the permanently randomized responses. Therefore, we can use the variance of the frequency estimation of each bit as a measure of utility. An advantage of the DP relaxation is that the frequency estimation process is straightforward and easy to implement. Since the probability distribution of the output $o_i$ after each round of relaxation is identical to that of an $\epsilon_i$-LDP randomized response, we can treat it as the corresponding randomized response and estimate the frequency of the original data using Eq. \eqref{eq:rr_frequency_estimate_binary}. The variance of the estimated frequency is

\begin{equation}
    Var(\hat{\pi}_b) = (\frac{e^{\epsilon_i} + 1}{ e^{\epsilon_i} - 1})^2  Var(\lambda_b).
\end{equation}

\noindent where  $\epsilon_i$ is the latest relaxed DP guarantee. Here, $\lambda_b = \frac{1}{N} \sum_j Y_j$, where $N$ is the number of clients and $Y_j$ is the relaxation result $o_i$ from the $j$-th client. The variance of $Y_j$ is identical to that of a Bernoulli distribution with a probability of $\frac{e^{\epsilon_i}}{e^{\epsilon_i} + 1}$, so we have

\begin{equation}
    Var(\hat{\pi}_b) = (\frac{e^{\epsilon_i} + 1}{ e^{\epsilon_i} - 1})^2 \frac{e^{\epsilon_1}}{N (e^{\epsilon_1} + 1)^2} = \frac{e^{\epsilon_i}}{N ( 1 - e^{\epsilon_i})^2}.
    \label{eq:var_binary_rr_f}
\end{equation}

Now, let's also derive the variance of the decoding process of the noisy samplings. Assume each client has sent $K$ noisy-sampled strings to the server. The decoding process can be split into two steps (again, we focus on one bit): \newline

1. Use $K$ reported noisy-sampled bits of a client to estimate the actual value of the perturbed bit $B'$ of the permanent randomized response of the client. \newline

2. Use the estimated $B'$ from all $N$ clients to estimate the frequency of the original bit $B$ being one.  \newline

The variance of the estimated frequency of $B = 1$ is derived as

\begin{equation}
    Var(\hat{B}) = \frac{\beta (1 - \beta)}{NK (1 - 2 \beta)^2 (1 - 2 \alpha)^2} + \frac{\alpha (1 - \alpha)}{N(1 - 2 \alpha)^2} .
    \label{eq:noisy_sampling_var_final}
\end{equation}

\noindent The proof is presented in Appendix \ref{sec:var_noisy_sampling}.

Fig. \ref{fig:var_rappor_full} compares the variance of the DP relaxation process and repeated noisy sampling. The repeated noisy sampling has $\epsilon_\alpha = 1$, and $\epsilon_\beta$ ranges from 0.1 to 1.0, while the DP relaxation process relaxes the DP guarantee in the same way as the repeated noisy sampling. Among different $\epsilon_\beta$, the DP relaxation outperforms the repeated noisy sampling with a much faster reduction in variance. Notably, \cite{huang2021instance} has proved that the frequency estimation, as a special case of mean estimation, from randomized responses has optimal utility. Thus, the frequency estimation from the outputs of the DP relaxation is also optimal.

Besides relaxing DP guarantee following Eq. \eqref{eq:eps_noisy_sampling}, we can relax the DP guarantee of the permanent randomized response of RAPPOR in our own preferred way. Fig. \ref{fig:var_rappor_linear} is an example of relaxing the DP guarantee linearly, which presents the corresponding variance of the frequency estimation of the original bit $B$. In the figure, all the cases start with an initial DP guarantee $\epsilon = 0.1$ and gradually relax $\epsilon$ to 1.0 at different rates.

In conclusion, we recommend replacing Steps 3 and 4 of RAPPOR with the DP relaxation, which will lead to the following benefits:

1. Optimal utility when decoding the collected data.

2. Privacy loss is more controllable. Using the DP relaxation, we can control exactly how to relax the DP guarantee. 

3. Minimum use of privacy budget. When the accuracy of the estimated frequency satisfies the goal, the relaxation can be terminated immediately.

\subsection{Mean Estimation}

Randomized responses can also be employed to estimate the mean of a group of continuous values \cite{nguyen2016collecting, ye2019privkv}. Let us denote an object in the group as $X$. The function $f$ returns a scalar such that for all $X$, $f(X)$ lies within the interval $[l, h]$, where $l$ and $h$ are constants. For instance, $f(x) = x$ and $f(x) = x^2$ are used to estimate $E[X]$ and $E[X^2]$, respectively. The differentially private mean estimation of $f(X)$ for a group can be calculated as follows \cite{nguyen2016collecting}:  \newline

1. Discretize $f(X)$ into either $l$ or $h$ with the following probabilities:

\begin{equation}
    Pr(z | X) = 
    \begin{cases}
        \frac{f(X) - l}{h - l} & z = h \\
        \frac{h - f(X)}{h - l} & z = l
    \end{cases} ,
\end{equation}

2. Apply the randomized response against $z$. \newline

Similarly to frequency estimation, after receiving a group of responses, the mean estimation can be derived as

\begin{equation}
    \hat{f} \approx l + h \frac{(e^\epsilon + 1) \lambda_h - 1}{e^\epsilon - 1} ,
    \label{eq:mean_estimation}
\end{equation}

\noindent where $\hat{f}$ is the mean estimator of $f(X)$ for the group, $\lambda_h$ is the frequency of the responses from Step 2 being $h$, and $\epsilon$ is the differential privacy guarantee of Step 2.

Therefore, DP relaxation can be applied to the randomized response of Step 2, assuming the $z$ value from Step 1 is stored and unchanged. The outputs after each relaxation can be placed into Eq. \eqref{eq:mean_estimation}, acquiring the updated mean estimation with respect to the relaxed DP guarantee.

\subsection{Differentially-private Data Market}

\begin{figure*}[ht!]
\centering
\subfloat[]{\label{fig:experiment_1_res} \includegraphics[width=0.32\textwidth]{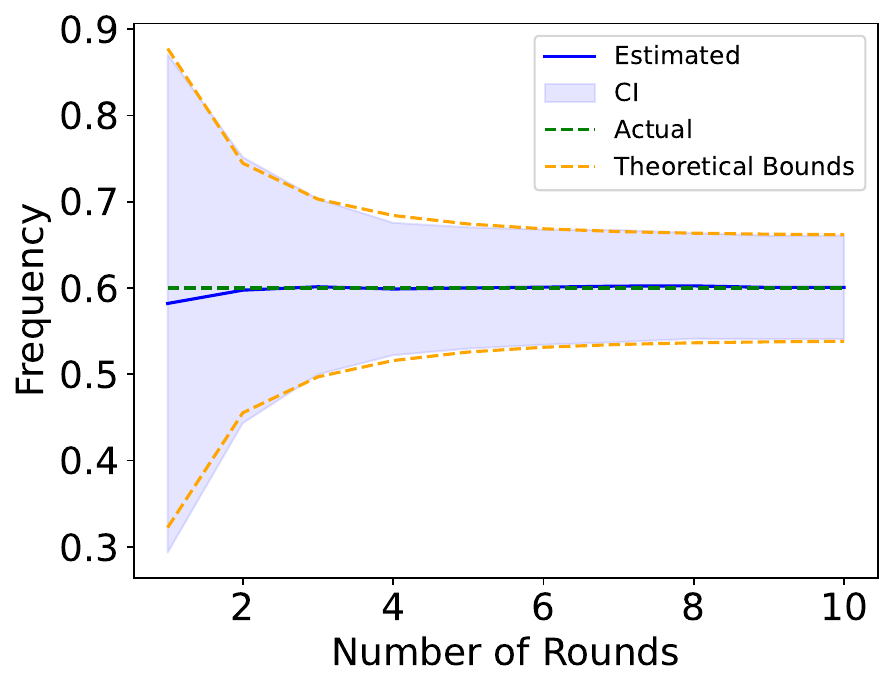}}
\hfill
\subfloat[]{\label{fig:experiment_1_var} \includegraphics[width=0.32\textwidth]{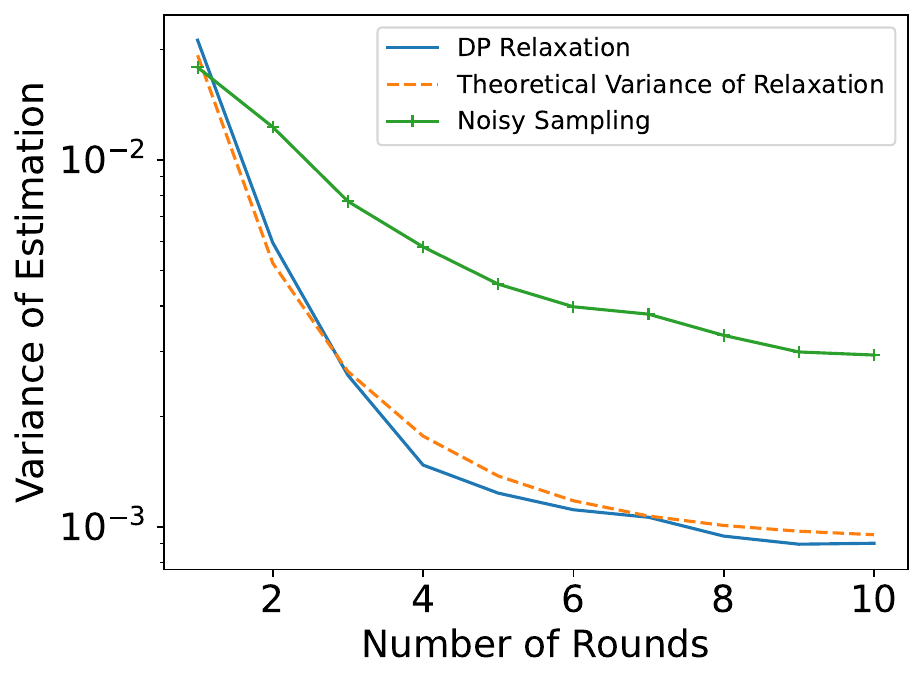}}%
\hfill
\subfloat[]{\label{fig:experiment_1_err} \includegraphics[width=0.32\textwidth]{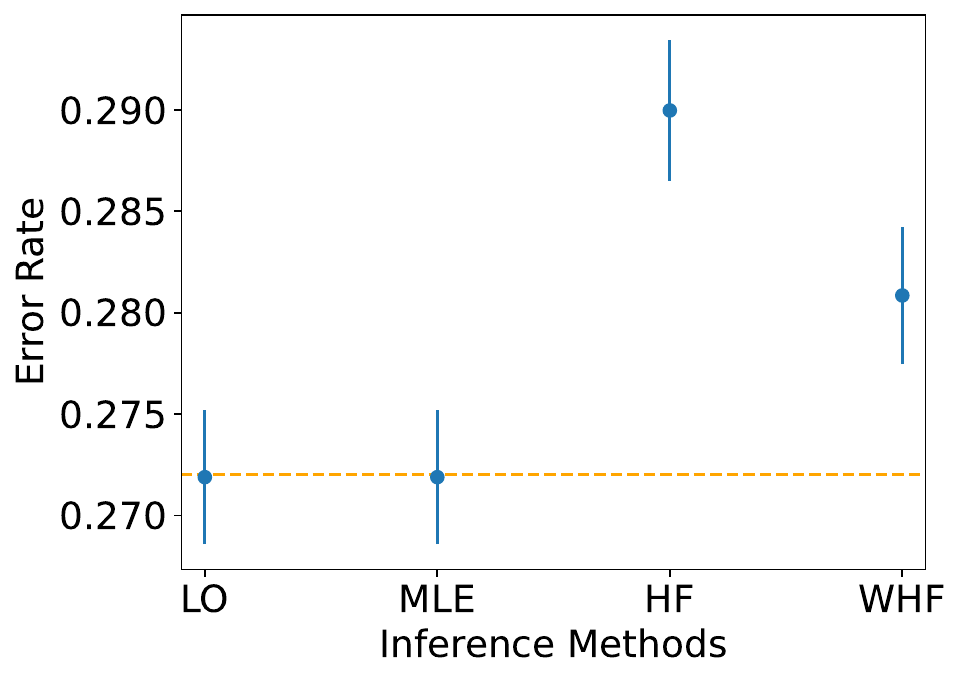}}%
\caption{(a) displays the mean and confidence interval ($2 \sigma$) of the estimated frequency of the objects being one in the first experiment. Dashed curves represent the theoretical values; (b) illustrates the variance of the estimated frequency for both the DP relaxation of randomized responses and repeated noisy sampling with the same DP guarantee. The dashed curve represents the theoretical variance of DP relaxation. (c) shows the mean and error bars of the error rates for different inference methods inferring the true values of the objects in the first experiment. The dashed line represents the theoretical minimum. LO, MLE, HF, and WHF stand for Last Output, Maximum Likelihood Estimation, Highest Frequency, and Weighted Highest Frequency, respectively.}
\end{figure*}

\begin{figure*}[ht!]
\centering
\subfloat[]{\label{fig:experiment_2_1} \includegraphics[width=0.32\textwidth]{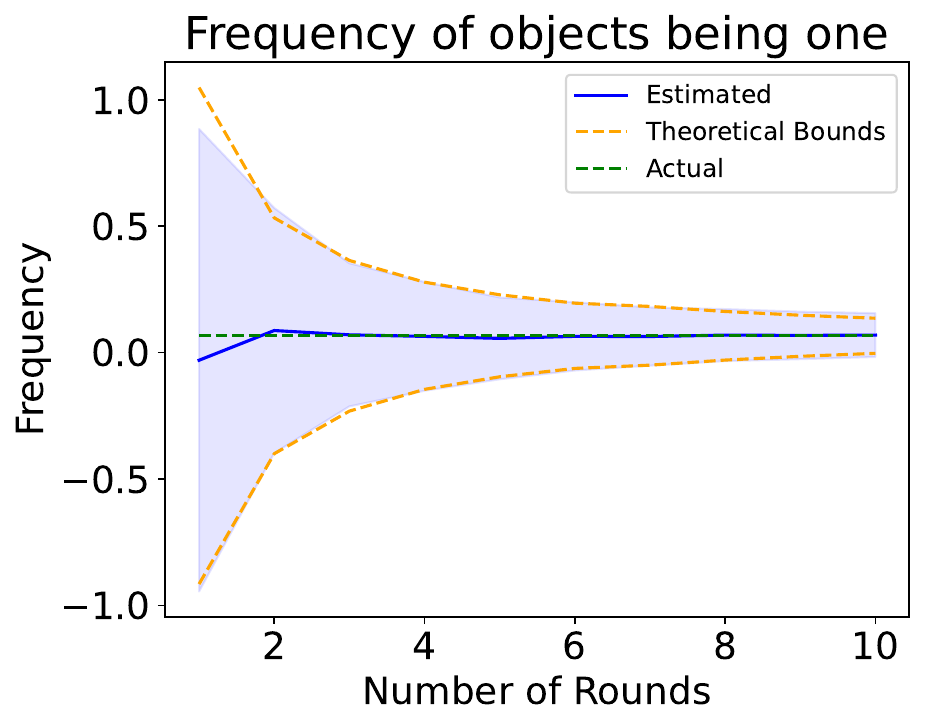}}
\hfill
\subfloat[]{\label{fig:experiment_2_2} \includegraphics[width=0.32\textwidth]{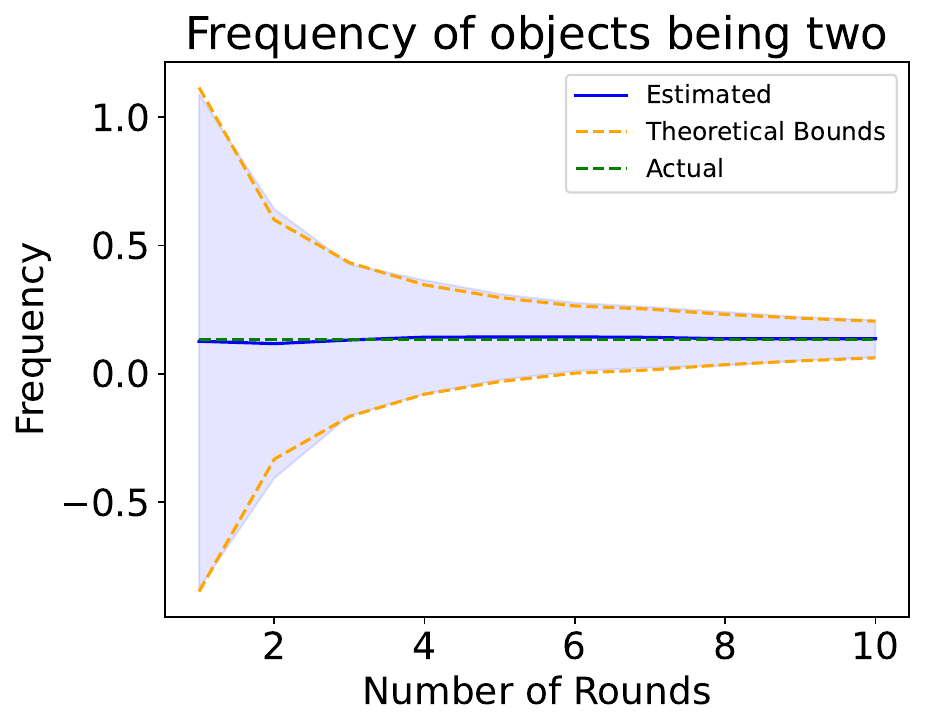}}%
\hfill
\subfloat[]{\label{fig:experiment_2_3} \includegraphics[width=0.32\textwidth]{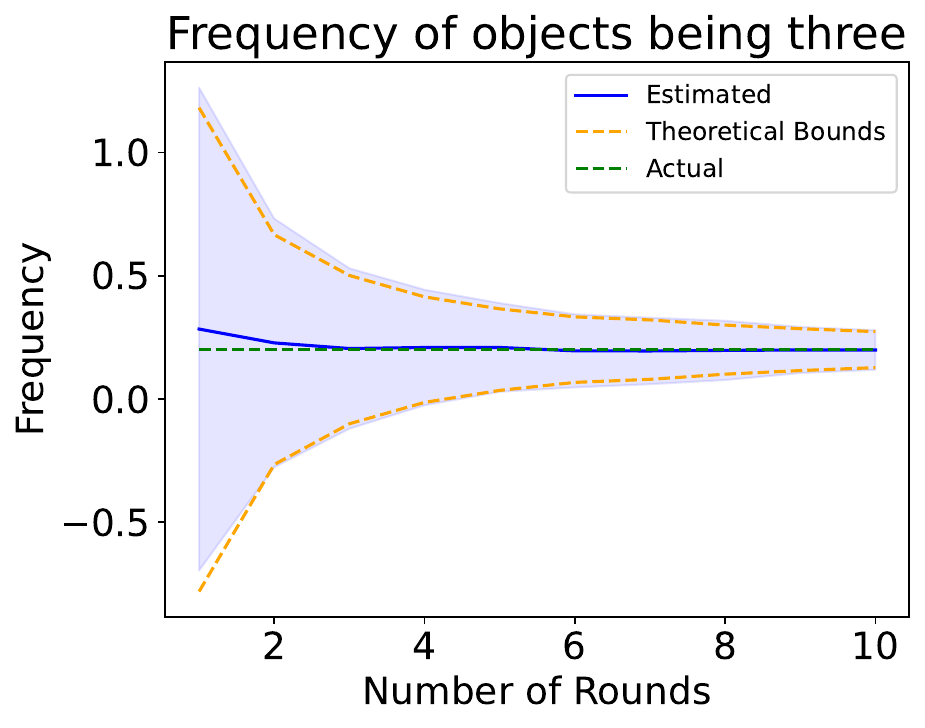}}%

\subfloat[]{\label{fig:experiment_2_4} \includegraphics[width=0.32\textwidth]{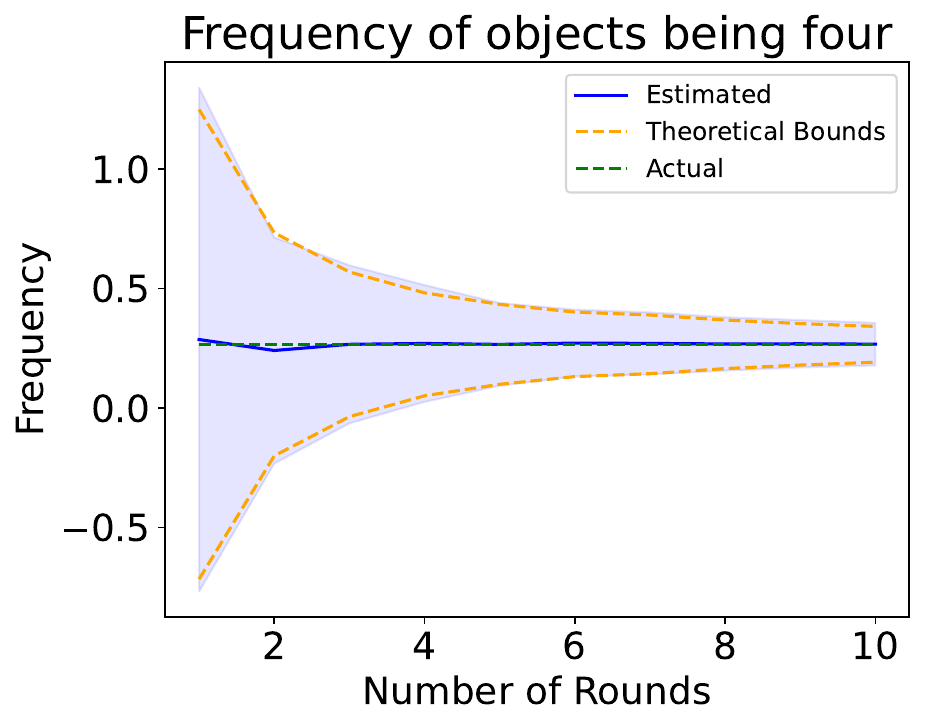}}
\hfill
\subfloat[]{\label{fig:experiment_2_5} \includegraphics[width=0.32\textwidth]{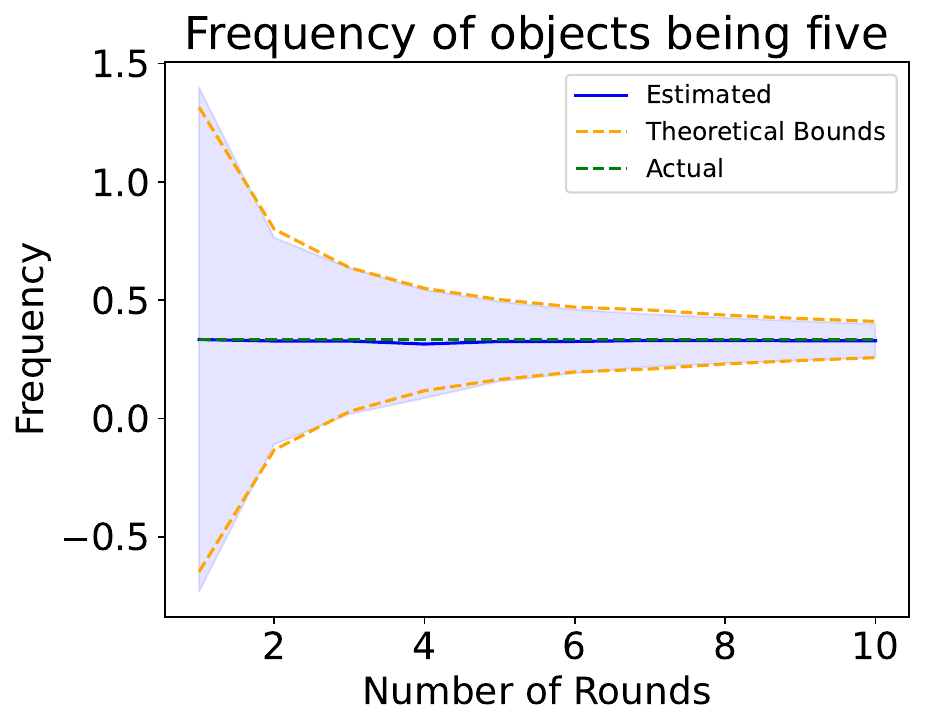}}%
\hfill
\subfloat[]{\label{fig:experiment_2_err} \includegraphics[width=0.32\textwidth]{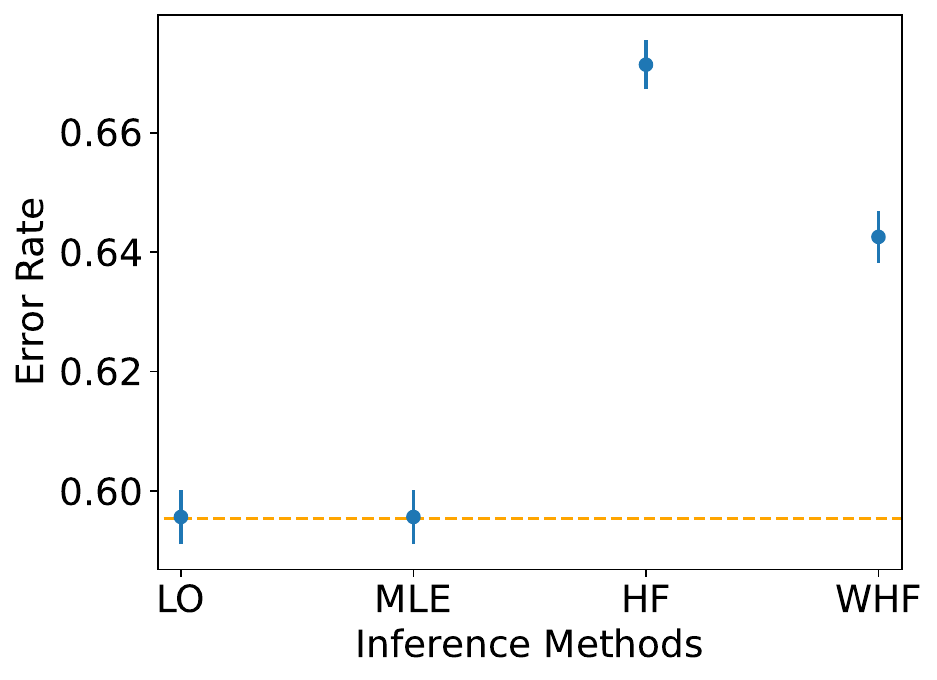}}%

\caption{(a) - (e) present the mean and confidence interval ($2 \sigma$) of the estimated frequency of the object in the second experiment being one to five, respectively. Dash curves represent the theoretical values; (f) depicts the mean and error bars of the error rates for different inference methods inferring the true values of the objects in the second experiment. The dashed line represents the theoretical minimum. LO, MLE, HF, and WHF stand for Last Output, Maximum Likelihood Estimation, Highest Frequency, and Weighted Highest Frequency, respectively.}
\end{figure*}

A differentially-private data market functions as a platform where users can sell their private data while maintaining the integrity of their privacy budget. The data available in the market is safeguarded by differential privacy, and data with a higher differential privacy guarantee holds greater value, as it is more likely to accurately represent the true value of the original data. The DP relaxation of randomized response enables users to sell and resell their data to multiple users with varying differential privacy guarantees, all while preserving the overall privacy budget.

For instance, a user may gradually relax the DP guarantee of the randomized response applied to their data from 0.1 to 1.0, using a small stride (e.g., 0.01), and store all the outputs generated during this relaxation process. A buyer has the flexibility to purchase the data with any desired DP guarantee, and they can revisit later to obtain the same data with a higher guarantee. Furthermore, the user can sell this data to multiple buyers, each with their preferred DP guarantee, and the overall privacy loss remains capped at $\epsilon = 1.0$.

\section{Experimental Evaluation}
\label{sec:experiment}

We will conduct two experiments to validate the correctness and utility of the DP relaxation of randomized response.

\subsection{Binary Randomized Response}

The first experiment focuses on the relaxation of binary randomized responses applied to 1000 objects with values of 600 ones and 400 zeros. The DP guarantee of the randomized response is relaxed following Eq. \eqref{eq:eps_noisy_sampling} with $\epsilon_\alpha = 1$ and $\epsilon_\beta= 0.5$, and it will undergo relaxation for 10 rounds. The randomized responses will be utilized to estimate the frequency of objects being one using Eq. \eqref{eq:rr_frequency_estimate_binary}. This experiment has been conducted 100 times. The mean and variance of the estimated frequency are plotted in Fig. \ref{fig:experiment_1_res}. The theoretical mean (0.6) and confidence interval ($2 \sigma$ from Eq. \eqref{eq:var_binary_rr_f}) are plotted for reference. The results perfectly match the theoretical values, indicating that the probability distribution of the relaxed outputs is consistent with randomized responses with the corresponding DP guarantee.

We also numerically compare the utility (variance) of the DP relaxation and repeated noisy samplings of RAPPOR. The noisy samplings are set up with the same parameters ($\epsilon_\alpha = 1$ and $\epsilon_\beta= 0.5$) and sampled 10 times as well. The variance of the estimated frequency of both DP relaxation and noisy samplings is plotted in Fig. \ref{fig:experiment_1_var}. The variance of the DP relaxation matches the theoretical value and outperforms that of the noisy samplings.

Additionally, we aim to validate that the DP relaxation complies with the DP guarantee. After 10 rounds of relaxation, it should protect data as well as the $\epsilon_{ns}(10)$-LDP randomized response, where $\epsilon_{ns}(10)$ is the value of Eq. \eqref{eq:eps_noisy_sampling} after 10 rounds. The optimality of Bayesian inference \cite{hutter2007universal} implies the following theorem:

\begin{theorem}
    If $Y$ is the result of an $\epsilon$-LDP randomized response against an object $X$ in a domain of size $m$, and the value of $X$ is uniformly sampled from $[m]$, then the error rate of inferring the value of $X$ given $Y$ without other auxiliary information is at least $1 - \frac{e^{\epsilon}}{e^{\epsilon} + m - 1}$.
    \label{thm:error_rate}
\end{theorem}

\noindent The error rate $1 - \frac{e^{\epsilon}}{e^{\epsilon} + m - 1}$ is derived from Bayesian inference with a uniform prior distribution (i.e., no auxiliary information besides $X$ being uniformly sampled). Similar theorems have been proven in existing literature, such as \cite{song2019common}. Therefore, we will employ several inference methods to guess the values of the objects in the experiment and examine if their error rates are at least $1 - \frac{e^{\epsilon_{ns}(10)}}{e^{\epsilon_{ns}(10)} + m - 1}$. The first method guesses the true value of an object as the last output of the relaxation, referred to as the \textit{Last Output} method. The second method is \textit{Max-Likelihood Estimation}, which computes the likelihood of the sequence of relaxed outputs from an object given different possible true values and guesses the true value to be the one with the maximum likelihood. The third method is \textit{Highest Frequency}, which counts the number of values (i.e., 0s and 1s in our case) in the sequence of relaxed outputs and guesses the true value to be the one with the largest count. The last method is \textit{Weighted Highest Frequency}, which sums the weights of the values (i.e., 0s and 1s in our case) in the sequence of relaxed outputs, and the weight of a value equals the corresponding DP guarantee of its output. Since Theorem \ref{thm:error_rate} requires ``$X$ is uniformly sampled from $[m]$", we will calculate the error rates among 400 of the objects with value one and all the objects with value zero. Given that the experiment is run 100 times, we can calculate the mean and standard deviation of the error rates of different methods, which are plotted in Fig. \ref{fig:experiment_1_err}. We observe that no inference methods violate the theoretical minimum, and \textit{Last Output} and \textit{Max-Likelihood Estimation} are the optimal inference methods because they are identical to Bayesian inference with uniform prior knowledge.

\subsection{Polychotomous Randomized Response}

The second experiment focuses on polychotomous randomized responses, involving 1500 objects with the values of 100 ones, 200 twos, 300 threes, 400 fours, and 500 fives. Randomized responses are applied to these objects to perturb their true values. The initial DP guarantee of the randomized responses is 0.1 and is linearly relaxed to 1.0 with a stride of 0.1. The randomized responses are utilized to estimate the frequency of different values (one to five) using Eq. \eqref{eq:rr_frequency_estimate_poly}. This experiment is also conducted 100 times. The mean and confidence interval of the estimated frequency of different values are plotted in Fig. \ref{fig:experiment_2_1} - \ref{fig:experiment_2_5}. The theoretical confidence interval of the estimated frequency is computed (see Appendix \ref{sec:var_estimated_f}) and plotted for reference, perfectly matching our experimental values. This indicates that the output from DP relaxation has the same probability distribution as the corresponding randomized response.

Additionally, we validate the DP guarantee of the relaxation by examining its error rates. Similar to the first experiment, we plot the error rates of different inference methods in Fig. \ref{fig:experiment_2_err}, and none of them are smaller than the theoretical minimum, implying the compliance of the DP guarantee of our relaxation algorithm.

\bibliographystyle{IEEEtran}
\bibliography{references}  
%

\appendices

\begin{table}[ht]
\centering
\begin{tabular}{| c | c c c c c|}
\hline
\backslashbox{m}{$\epsilon_i$} & 0.1 & 0.5 & 1.0 & 2.0 & 10\\
\hline
3 & & 0.584 & 0.840 & 0.943 & 1.000 \\
4 & & 0.511 & 0.802 & 0.922 & 1.000 \\
5 & & 0.463 & 0.775 & 0.906 & 1.000 \\
6 & & 0.430 & 0.755 & 0.891 & 1.000 \\
7 & & 0.405 & 0.740 & 0.879 & 1.000 \\
8 & & 0.386 & 0.728 & 0.869 & 1.000 \\
9 & & 0.371 & 0.718 & 0.860 & 1.000 \\
10 & & 0.359 & 0.710 & 0.852 & 1.000 \\
\hline
\end{tabular}

\vspace{5pt}
{\raggedright The value at $\epsilon_i$ represents the relaxation from $\epsilon_{i-1}$ to $\epsilon_i$. Same for the tables below. \par}
\caption{$Pr(o_2 = a | o_1 = a, X = a)$ of the DP relaxation of randomized response.}
\label{tab:p_aa_example}
\vspace{10pt}

\begin{tabular}{| c | c c c c c|}
\hline
\backslashbox{m}{$\epsilon_i$} & 0.1 & 0.5 & 1.0 & 2.0 & 10\\
\hline
3 & & 0.392 & 0.509 & 0.347 & 0.000 \\
4 & & 0.342 & 0.486 & 0.339 & 0.000 \\
5 & & 0.310 & 0.470 & 0.333 & 0.000 \\
6 & & 0.288 & 0.458 & 0.328 & 0.000 \\
7 & & 0.272 & 0.449 & 0.324 & 0.000 \\
8 & & 0.259 & 0.442 & 0.320 & 0.000 \\
9 & & 0.249 & 0.436 & 0.316 & 0.000 \\
10 & & 0.241 & 0.431 & 0.314 & 0.000 \\
\hline
\end{tabular}
\caption{$Pr(o_2 = b | o_1 = b, X = a)$ of the DP relaxation of randomized response.}
\vspace{10pt}

\begin{tabular}{| c | c c c c c|}
\hline
\backslashbox{m}{$\epsilon_i$} & 0.1 & 0.5 & 1.0 & 2.0 & 10\\
\hline
3 & & 0.379 & 0.359 & 0.575 & 1.000 \\
4 & & 0.297 & 0.296 & 0.520 & 1.000 \\
5 & & 0.245 & 0.252 & 0.474 & 1.000 \\
6 & & 0.208 & 0.219 & 0.436 & 0.999 \\
7 & & 0.181 & 0.194 & 0.403 & 0.999 \\
8 & & 0.160 & 0.174 & 0.375 & 0.999 \\
9 & & 0.143 & 0.158 & 0.351 & 0.999 \\
10 & & 0.130 & 0.144 & 0.330 & 0.999 \\
\hline
\end{tabular}
\caption{$Pr(o_2 = a | o_1 = b, X = a)$ of the DP relaxation of randomized response.}
\label{tab:p_ba_example}
\end{table}

\section{Proof of LDP Requirement for the Relaxation of Binary Randomized Response}
\label{sec:proof_ldp_binary}

We will demonstrate that Eq. \eqref{eq:ldp_composition_binary_sufficient_step_1} is sufficient to satisfy Property \ref{pro:ldp_composition_binary}. It is trivial to see that when $o_1 = a$ and $o_2 = a$, Eq. \eqref{eq:ldp_composition_binary_sufficient_step_1} is equivalent to Eq. \eqref{eq:ldp_composition_binary} of Property \ref{pro:ldp_composition_binary}. 

In the case where $o_1 = a$ and $o_2 = b$, considering symmetry between $a$ and $b$, we have $Pr(o_1 = a, o_2 = b | X = b) = Pr(o_1 = b, o_2 = a | X = a)$, so Eq. \eqref{eq:ldp_composition_binary} in this scenario can be rewritten as
\begin{multline}
    Pr(o_1 = a, o_2 = b | X = a) \\ \le e^{\epsilon_2} Pr(o_1 = b, o_2 = a | X = a) ,
\end{multline}

\noindent which can be transformed as 
\begin{multline}
    Pr(o_1 = a | X = a) (1 - p_a) \\ \le e^{\epsilon_2}  Pr(o_1 = b | X = a) (1 - p_b) .
\end{multline}

\noindent Substituting $Pr(o_1 = a | X = a) = \frac{e^{\epsilon_1}}{e^{\epsilon_1} + 1}$  and $Pr(o_1 = b | X = a) = \frac{1}{e^{\epsilon_1} + 1}$, we have

\begin{equation}
    e^{\epsilon_1} (1 - p_a) \\ \le e^{\epsilon_2} (1 - p_b) .
\end{equation}

\noindent Substituting $p_a$ and $p_b$, the above inequality becomes

\begin{equation}
    e^{\epsilon_1} \frac{e^{-\epsilon_1} - e^{-\epsilon_2}}{e^{\epsilon_2} - e^{-\epsilon_2}} \le e^{\epsilon_2} \frac{e^{2\epsilon_2} - e^{\epsilon_1 + \epsilon_2}}{e^{2\epsilon_2} - 1} ,
\end{equation}

\noindent which can be simplified as 

\begin{equation}
    1 \le e^{2 \epsilon_2} .
    \label{eq:ldp_composition_binary_ab_final}
\end{equation}

\noindent Therefore, to prove Eq. \eqref{eq:ldp_composition_binary} with $o_1 = a$ and $o_2 = b$, we need to prove Eq. \eqref{eq:ldp_composition_binary_ab_final}, which is always true because $\epsilon_2 > 0$. Thus, Eq. \eqref{eq:ldp_composition_binary} in this case is proved.

When $o_1 = b$ and $o_2 = a$, after considering the symmetry $Pr(o_1 = b, o_2 = a | X = b) = Pr(o_1 = a, o_2 = b | X = a)$, we have Eq. \eqref{eq:ldp_composition_binary} to be rewritten as
\begin{multline}
    Pr(o_1 = b, o_2 = a | X = a) \\ \le e^{\epsilon_2} Pr(o_1 = a, o_2 = b | X = a) .
\end{multline}

\noindent Similar to above, when substituting $Pr(o_1 | X)$, $p_a$ and $p_b$, we have

\begin{equation}
     (1 - p_b) \\ \le e^{\epsilon_2} e^{\epsilon_1} (1 - p_a) .
\end{equation}

\noindent Substituting $p_a$ and $p_b$, the above inequality becomes

\begin{equation}
     \frac{e^{2\epsilon_2} - e^{\epsilon_1 + \epsilon_2}}{e^{2\epsilon_2} - 1}  \le e^{\epsilon_1 + \epsilon_2} \frac{e^{-\epsilon_1} - e^{-\epsilon_2}}{e^{\epsilon_2} - e^{-\epsilon_2}}  ,
\end{equation}

\noindent which can be simplified as $1 \le 1$. Therefore, Eq. \eqref{eq:ldp_composition_binary} with $o_1 = b$ and $o_2 = a$ is proved.

Finally, when $o_1 = b$ and $o_2 = b$, after considering the symmetry $Pr(o_1 = b, o_2 = b | X = b) = Pr(o_1 = a, o_2 = a | X = a)$, we have Eq. \eqref{eq:ldp_composition_binary} to be rewritten as
\begin{multline}
    Pr(o_1 = b, o_2 = b | X = a) \\ \le e^{\epsilon_2} Pr(o_1 = a, o_2 = a | X = a) .
\end{multline}

\noindent Similar to above, when substituting $Pr(o_1 | X)$, $p_a$ and $p_b$, we have

\begin{equation}
     p_b \\ \le e^{\epsilon_2} p_a .
     \label{eq:ldp_composition_binary_bb_final}
\end{equation}

\noindent Since $p_a = e^{\epsilon_2 - \epsilon_1} p_b \le p_b$ and $e^{\epsilon_1} \le 1$, Eq. \eqref{eq:ldp_composition_binary_bb_final} is always satisfied, so Eq. \eqref{eq:ldp_composition_binary} with $o_1 = b$ and $o_2 = a$ is proved to be valid. $\qed$

\section{Detailed Derivation of LDP Requirement of Relaxation}
\label{sec:detailed_ldp_proof}

This section will derive Inequalities \eqref{eq:ab_ba} - \eqref{eq:cb_ca} from Property \ref{pro:ldp_composition_poly}:

When $o_1 = a$ and $o_2 = b$, we have
\begin{multline}
    Pr(o_1 = a, o_2 = b | X = a) \\ \le e^{\epsilon_2} Pr(o_1 = a, o_2 = b | X = b) .
\end{multline}

\noindent Given the symmetry we define for the conditional probability, we have $Pr(o_1 = a, o_2 = b | X = b) = Pr(o_1 = b, o_2 = a | X = a)$, so the above equation can be rewritten as
\begin{multline}
    Pr(o_1 = a, o_2 = b | X = a) \\ \le e^{\epsilon_2} Pr(o_1 = b, o_2 = a | X = a) .
\end{multline}

\noindent That is equivalent to 

\begin{equation}
    Pr(o_1 = a | X = a) \frac{1 - p_{aa}}{m - 1} \le Pr(o_1 = b | X = a) e^{\epsilon_2} p_{ba}.
\end{equation}

\noindent Given that $o_1$ is generated by $\epsilon_1$-LDP randomized response, we have $Pr(o_1 = a | X = a) = \frac{e^{\epsilon_1}}{e^{\epsilon_1} + m - 1}$ and $Pr(o_1 = b | X = a) = \frac{1}{e^{\epsilon_1} + m - 1}$ where $b \ne a$. Substituting them into the above equation, we have

\begin{equation}
    e^{\epsilon_1} \frac{1 - p_{aa}}{m - 1} \le e^{\epsilon_2} p_{ba} ,
\end{equation}

\noindent which is Eq. \eqref{eq:ab_ba}. Given the symmetry of LDP definition, we also require:
\begin{multline}
    Pr(o_1 = a, o_2 = b | X = b) \\ \le e^{\epsilon_2} Pr(o_1 = a, o_2 = b | X = a) ,
\end{multline}

\noindent which is equivalent to
\begin{multline}
    Pr(o_1 = b, o_2 = a | X = a) \\ \le e^{\epsilon_2} Pr(o_1 = b, o_2 = a | X = b) ,
\end{multline}

\noindent and can be simplified as

\begin{equation}
     p_{ba} \le e^{\epsilon_2} e^{\epsilon_1} \frac{1 - p_{aa}}{m - 1}
\end{equation}

\noindent in a similar way. This inequality is Eq. \eqref{eq:ba_ab}.

When $o_1 = a$ and $o_2 = c$, we have

\begin{multline}
    Pr(o_1 = a, o_2 = c | X = a) \\ \le e^{\epsilon_2} Pr(o_1 = a, o_2 = c | X = b) .
\end{multline}

\noindent Given the symmetry, we have $Pr(o_1 = a, o_2 = c | X = b) = Pr(o_1 = b, o_2 = c | X = a)$, so the above equation can be rewritten as
\begin{multline}
    Pr(o_1 = a, o_2 = c | X = a) \\ \le e^{\epsilon_2} Pr(o_1 = b, o_2 = c | X = a) .
\end{multline}

\noindent which is equivalent to 

\begin{equation}
    Pr(o_1 = a | X = a) \frac{1 - p_{aa}}{m - 1} \le e^{\epsilon_2} \frac{1 - p_{ba} - p_{bb}}{m - 2}.
\end{equation}

\noindent Similar to above, Substituting $Pr(o_1 | X = a)$ into the above equation, we have

\begin{equation}
    e^{\epsilon_1} \frac{1 - p_{aa}}{m - 1} \le e^{\epsilon_2} \frac{1 - p_{ba} - p_{bb}}{m - 2},
\end{equation}

\noindent which is Eq. \eqref{eq:ac_bc}. Given the symmetry of LDP definition, we also require:
\begin{multline}
     Pr(o_1 = a, o_2 = c | X = b) \\ \le e^{\epsilon_2} Pr(o_1 = a, o_2 = c | X = a),
\end{multline}

\noindent which is equivalent to 

\begin{multline}
     Pr(o_1 = b, o_2 = c | X = a) \\ \le e^{\epsilon_2} Pr(o_1 = b, o_2 = c | X = b),
\end{multline}

\noindent and can be further simplified as

\begin{equation}
     \frac{1 - p_{ba} - p_{bb}}{m - 2}  \le e^{\epsilon_2} e^{\epsilon_1} \frac{1 - p_{aa}}{m - 1},
\end{equation}

\noindent in a similar way. This inequality is Eq. \eqref{eq:bc_ac}.

When $o_1 = c$ and $o_2 = a$, we have

\begin{multline}
    Pr(o_1 = c, o_2 = a | X = a) \\ \le e^{\epsilon_2} Pr(o_1 = c, o_2 = a | X = b) .
\end{multline}

\noindent Given the symmetry, we have $Pr(o_1 = c, o_2 = a | X = b) = Pr(o_1 = c, o_2 = b | X = a)$, so the above equation can be rewritten as
\begin{multline}
    Pr(o_1 = c, o_2 = a | X = a) \\ \le e^{\epsilon_2} Pr(o_1 = c, o_2 = b | X = a) .
\end{multline}

\noindent which is equivalent to 

\begin{equation}
    Pr(o_1 = c | X = a) p_{ba} \le Pr(o_1 = c | X = a) e^{\epsilon_2} \frac{1 - p_{ba} - p_{bb}}{m - 2}.
\end{equation}

\noindent Canceling $Pr(o_1 = c | X = a)$, we have

\begin{equation}
     p_{ba} \le e^{\epsilon_2} \frac{1 - p_{ba} - p_{bb}}{m - 2}.
\end{equation}

\noindent which is Eq. \eqref{eq:ca_cb}. Given the symmetry of LDP definition, we also require:
\begin{multline}
     Pr(o_1 = c, o_2 = a | X = b) \\ \le e^{\epsilon_2} Pr(o_1 = c, o_2 = a | X = a),
\end{multline}

\noindent which is equivalent to 

\begin{multline}
     Pr(o_1 = c, o_2 = b | X = a) \\ \le e^{\epsilon_2} Pr(o_1 = c, o_2 = b | X = b),
\end{multline}

\noindent and can be further simplified as

\begin{equation}
     \frac{1 - p_{ba} - p_{bb}}{m - 2}  \le e^{\epsilon_2} p_{ba}
\end{equation}

\noindent in a similar way. This inequality is Eq. \eqref{eq:cb_ca}.

$o_1$ and $o_2$ can be either $a$, $b$, or $c$, so there are 9 possible cases. Section \ref{sec:single_relax_poly} has proved three cases, and this section has proved the remaining six. Thus, all cases have been considered.

\section{Finding and Prove the Solution of Relaxing Polychotomous Randomized Response}
\label{sec:feasibility}

Equation \eqref{eq:end_state_e2} and Inequalities \eqref{eq:aa_bb}, \eqref{eq:bb_aa}, \eqref{eq:ab_ba} - \eqref{eq:cb_ca}, and \eqref{eq:basic_prob} collectively constitute a linear programming problem involving the variables $p_{aa}$, $p_{ba}$, and $p_{bb}$. Solving this linear programming is equivalent to determining the values of $p_{aa}$, $p_{ba}$, and $p_{bb}$, and the resulting numerical solution may provide insights into the analytical solution. We numerically solve this problem for all cases where $\epsilon_1$ and $\epsilon_2$ range from 0.1 to 2 with a stride of 0.1, and the domain size $m$ ranges from 3 to 10. We observe that all the cases mentioned above have only one solution.

In linear programming, an inequality is considered a support if its equality sign is consistently satisfied. If the number of supports in a linear-programming problem is at least as many as the variables, then the solution will be unique. We have determined that, among all the aforementioned cases, Inequalities \eqref{eq:aa_bb}, \eqref{eq:ba_ab}, and \eqref{eq:ca_cb} consistently support our linear-programming problem. Additionally, Eq. \eqref{eq:end_state_e2} serves as another support, giving us a total of four supports for three variables. Consequently, the solution is always unique, and one support can be derived from the other three.

Assuming the equality sign of Inequalities \eqref{eq:aa_bb}, \eqref{eq:ba_ab} is consistently satisfied and combining them with Eq. \eqref{eq:end_state_e2}, we have derived the analytical solution for $p_{aa}$, $p_{ba}$, and $p_{bb}$, expressed as Eq. \eqref{eq:solution_poly}.

Now, we will demonstrate that Eq. \eqref{eq:solution_poly} satisfies Eq. \eqref{eq:end_state_e2} and Inequalities \eqref{eq:aa_bb}, \eqref{eq:bb_aa}, \eqref{eq:ab_ba} - \eqref{eq:cb_ca}, and \eqref{eq:basic_prob}. The proof of Eq. \eqref{eq:end_state_e2}, \eqref{eq:aa_bb}, \eqref{eq:ba_ab} is trivial because the solution is derived from them, and we will skip them.

When substituting Eq. \eqref{eq:solution_poly} into Inequality \eqref{eq:bb_aa}, we have

\begin{equation}
    p_{bb} \le e^{\epsilon_2} e^{\epsilon_2} p_{bb} ,
\end{equation}

\noindent which is always true given $e^{\epsilon_2} \ge 1$. Thus, Eq. \eqref{eq:solution_poly} satisfies Inequality \eqref{eq:bb_aa}.

When substituting Eq. \eqref{eq:solution_poly} into Inequality \eqref{eq:ab_ba}, we have

\begin{equation}
    \frac{(e^{\epsilon_1} - e^{\epsilon_2}) (e^{\epsilon_2 + 1}) }{e^{\epsilon_2}  + m - 1} \le 0 ,
\end{equation}

\noindent which is always true given $e^{\epsilon_1} \le e^{\epsilon_2}$ and $m - 1 \ge 2$, so Eq. \eqref{eq:solution_poly} satisfies Inequality \eqref{eq:ab_ba}.

When substituting Eq. \eqref{eq:solution_poly} into Inequality \eqref{eq:ac_bc}, we have

\begin{equation}
    \frac{e^{\epsilon_1} - e^{\epsilon_2}}{e^{\epsilon_2}  + m - 1} \le 0 ,
\end{equation}

\noindent which is always true given $e^{\epsilon_1} \le e^{\epsilon_2}$ and $m - 1 \ge 2$, so Eq. \eqref{eq:solution_poly} satisfies Inequality \eqref{eq:ac_bc}.

When substituting Eq. \eqref{eq:solution_poly} into Inequality \eqref{eq:bc_ac}, we have

\begin{equation}
    \frac{e^{\epsilon_1} - e^{\epsilon_2}}{e^{\epsilon_2}  + m - 1} \le 0 ,
\end{equation}

\noindent which is the same as above, so Eq. \eqref{eq:solution_poly} satisfies Inequality \eqref{eq:bc_ac} too.

When substituting Eq. \eqref{eq:solution_poly} into \eqref{eq:ca_cb}, we find out both sides are crossed out, so the equality sign of Inequality \eqref{eq:ca_cb} is always satisfied. 

Subsequently, if we substitute Eq. \eqref{eq:ca_cb} (equality sign here) into Inequality \eqref{eq:cb_ca}, we have

\begin{equation}
    e^{-\epsilon_2} p_{ba} \le e^{2\epsilon_2} p_{ba} ,
\end{equation}

\noindent which is obviously true, so Eq. \eqref{eq:solution_poly} satisfies Inequality \eqref{eq:cb_ca} too.

Now, we will prove that Eq. \eqref{eq:solution_poly} satisfies Inequality \eqref{eq:basic_prob}. First, reformulate $p_{bb}$ as

\begin{equation}
    p_{bb} =\frac{e^{\epsilon_1} (e^{\epsilon_2} + m - 1) - (e^{\epsilon_1}  + m - 1)}{(e^{\epsilon_2} - 1)(e^{\epsilon_2} + m - 1)} .
\end{equation}

\noindent Given that $e^{\epsilon_1} \ge 1$, $e^{\epsilon_2}  + m - 1 \ge e^{\epsilon_1}  + m - 1$, and $m - 1 \ge 2$, we have $p_{bb} \ge 0$. Given that the equality sign of Inequality \eqref{eq:aa_bb} is always satisfied, we have $p_{aa} \ge 0$ too. 

For $p_{ba}$, its numerator can be formulated as 

\begin{equation}
    e^{\epsilon_2} (e^{\epsilon_2} - e^{\epsilon_1}) .
\end{equation}
Given that $\epsilon_2 \ge \epsilon_1$, we have  $p_{ba} \ge 0$.

Then, we formulate $p_{aa} - 1$ as 

\begin{equation}
    p_{aa} - 1 =\frac{ (m - 1) (1 - e^{\epsilon_2 - \epsilon_1}) }{(e^{\epsilon_2} - 1)(e^{\epsilon_2} + m - 1)} .
\end{equation}

\noindent Since $e^{\epsilon_2 - \epsilon_1} \ge 1$, $p_{aa} - 1 \le 0$ always holds true. Similarly, we formulate $p_{bb} + p_{ba} - 1$ as

\begin{equation}
    p_{bb} + p_{ba} - 1 =\frac{ (m - 2) ( e^{\epsilon_1} - e^{\epsilon_2}) }{(e^{\epsilon_2} - 1)(e^{\epsilon_2} + m - 1)} ,
\end{equation}

\noindent which always is always not greater than zero, because $ e^{\epsilon_1} - e^{\epsilon_2} \le 0$. Therefore, Eq. \eqref{eq:solution_poly} has been proved to satisfy Eq. \eqref{eq:end_state_e2} and Inequalities \eqref{eq:aa_bb}, \eqref{eq:bb_aa}, \eqref{eq:ab_ba} - \eqref{eq:cb_ca} and \eqref{eq:basic_prob}. $\qed$

\section{Collusion-Proofness Proof}
\label{sec:proof_collusion_proof}

Given that

\begin{equation}
    Q(X | y) = \frac{Pr(y | x) Q(X)}{\sum\limits_{x' \in [m]} Pr(y | x') Q(x')} ,
\end{equation}

\noindent where $y$ is any information, Eq. \eqref{eq:collusion_proof} can be expanded as
\begin{equation}
    \frac{Pr(o_{\le n} | X)}{\sum\limits_{x' \in [m]} Pr(o_{\le n} | x') Q(x')} = \frac{Pr(o_n | X)}{\sum\limits_{x' \in [m]} Pr(o_n | x') Q(x')} ,
\end{equation}

\noindent and organized as
\begin{equation}
    \frac{Pr(o_n | X)}{Pr(o_{\le n} | X)} = \frac{\sum\limits_{x' \in [m]} Pr(o_n | x') Q(x')}{\sum\limits_{x' \in [m]} Pr(o_{\le n} | x') Q(x')} .
    \label{eq:collusion_proof_step_2}
\end{equation}

Recall Lemma \ref{lem:xxo==o}, it is equivalent to say, for any value of $X$, we have

\begin{equation}
    \frac{Pr(o_n | X)}{Pr(o_{\le n} | X)} = R(o_{\le n}) .
\end{equation}

\noindent Substitute it into the right hand side of Eq. \eqref{eq:collusion_proof_step_2}, we have
\begin{multline}
    \text{R.H.S of \eqref{eq:collusion_proof_step_2}} = 
    \frac{\sum\limits_{x' \in [m]} R(o_{\le n}) Pr(o_{\le n} | x') Q(x')}{\sum\limits_{x' \in [m]}Pr(o_{\le n} | x') Q(x')} \\ = R(o_{\le n}) = \frac{Pr(o_n | X)}{Pr(o_{\le n} | X)} . \qed
\end{multline}

\section{DP Guarantee of Noisy Sampling in RAPPOR}
\label{sec:dp_noisy_sampling}

This section will derive the DP guarantee of RAPPOR after running Step 3 for $K$ times. Here is a recap of the problem: denote $B$ as the original bit. In Step 2 of RAPPOR, an $\epsilon_\alpha$-LDP randomized response is applied to $B$, and it outputs $B'$, with the condition $Pr(B' = B) = \alpha$. Step 3 applies another $\epsilon_\beta$-LDP randomized response to $B'$, resulting in output $S$ with the condition $Pr(S = B') = \beta$. Step 3 will be executed $K$ times, and the number of occurrences where $S$ is one is denoted as $N_{S=1}$, following a binomial distribution.

The DP guarantee of RAPPOR after running Step 3 for $K$ times can be derived from:
\begin{equation}
    \epsilon = \max\limits_{k} |\ln \frac{Pr(N_{S=1} = k | B = 1)}{Pr(N_{S=1} = k | B = 0)} |.
\end{equation}

\noindent Given the aforementioned distribution of $B'$ and $S$, we have
\begin{multline}
    Pr(N_{S=1} = k | B = 1) = \alpha C_{K}^k \beta^k (1 - \beta)^{K-k}  \\ + (1 - \alpha) C_{K}^k \beta^{K-k} (1 - \beta)^k 
\end{multline}

\noindent and
\begin{multline}
    Pr(N_{S=1} = k | B = 0) = \alpha  C_{K}^k \beta^{K-k} (1 - \beta)^k  \\ + (1 - \alpha) C_{K}^k \beta^k (1 - \beta)^{K-k} ,
\end{multline}

\noindent where $C_{K}^k$ represents "Choose $k$ from $K$". Assuming $\alpha > 0.5$ and $\beta > 0.5$, by taking the derivative of $\frac{Pr(N_{S=1} = k | B = 1)}{Pr(N_{S=1} = k | B = 0)}$, we have seen it increase with $k$ increasing. Thus, 
\begin{multline}
 \max\limits_{k} \frac{Pr(N_{S=1} = k | B = 1)}{Pr(N_{S=1} = k | B = 0)} \\ = \frac{Pr(N_{S=1} = K | B = 1)}{Pr(N_{S=1} = K | B = 0)}.
\end{multline}

\noindent and
\begin{multline}
 \min \limits_{k} \frac{Pr(N_{S=1} = k | B = 1)}{Pr(N_{S=1} = k | B = 0)} \\ = \frac{Pr(N_{S=1} = 0 | B = 1)}{Pr(N_{S=1} = 0 | B = 0)},
\end{multline}

\noindent which actually equals $\frac{Pr(N_{S=1} = K | B = 1)}{Pr(N_{S=1} = K | B = 0)}$ too. Therefore, the DP guarantee is derived as

\begin{equation}
    \epsilon = \ln \frac{\alpha \beta^K + (1 - \alpha) (1 - \beta)^{K}}{\alpha (1 - \beta)^{K} + (1 - \alpha) \beta^K} .
\end{equation}

If we substitute $e^{\epsilon_\alpha} = \frac{\alpha}{1 - \alpha}$ and $e^{\epsilon_\beta} = \frac{\beta}{1 - \beta}$ into the above equation, we have

\begin{equation}
    \epsilon = \ln \frac{e^{\epsilon_\alpha} e^{K \epsilon_\beta} + 1}{e^{\epsilon_\alpha} + e^{K \epsilon_\beta}} . \qed
\end{equation}

\section{Variance of Estimation from Noisy Samplings}
\label{sec:var_noisy_sampling}

This section will derive the variance of the estimated frequency of the original bit from Step 2 of RAPPOR being ones after the repeated noisy sampling has been run $K$ times. Please refer to Appendix \ref{sec:dp_noisy_sampling} for a recap of the problem.

The estimation of $B'$ is equivalent to estimating frequency of $B' = 1$ from randomized responses, which can be achieved by Eq. \eqref{eq:rr_frequency_estimate_binary}. Thus, the variance of the estimator $\hat{B}'$ is formulated as

\begin{equation}
    Var(\hat{B'}) = \frac{Var(\lambda_{S})}{ (1 - 2 \beta)^2} ,
    \label{eq:hat_B'}
\end{equation}

\noindent where $\lambda_{S}$ is the frequency of $S = 1$ from a client, which is defined as $\frac{\sum_{j \in [K]} S_j}{K}$, where $S_j$ is the $j$-th reported bit. 

Now we will derive $Var(\lambda_{S})$. Without loss of generality, assume the original bit $B = 1$, and $B'$ is equivalent to a random variable with a Bernoulli distribution where $Pr(B' = 1) = \alpha$. If $B' = 1$, $\sum_{j \in [k]} S_j$ (number of $S$ being one) is a binomial distribution with probability of $\beta$ generating ones, which is denoted as $Y$. Similarly, if $B' = 0$, $\sum_{j \in [k]} S_j$ is a binomial distribution with probability of $1 - \beta$ generating ones, which is denoted as $Z$. Thus, the number of $S = 1$ considering both Steps 2 and 3, denoted as $\mathbf{S}$, can be represented as

\begin{equation}
    \mathbf{S} = B Y + (1 - B) Z .
\end{equation}

\noindent Therefore, the variance of $\mathbf{S}$ can be derived as 

\begin{multline}
    Var(\mathbf{S}) = Var(BY) + Var((1 - B) Z) \\ + 2 Cov(BY, (1 - B) Z).
\end{multline}

\noindent Given that $B$ and $Y$ are independent, and $Var(B) = \alpha (1 - \alpha)$, $E[B] = \alpha$, $Var(Y) = K \beta (1 - \beta)$, and $E[Y] = K \beta$, we have
\begin{multline}
    Var(BY) \\ = Var(B) Var(Y) + Var(B) E[Y]^2 + Var(Y) E[B]^2 \\ = \alpha (1 - \alpha) \beta (1 - \beta) K + \alpha (1 - \alpha)  (K \beta)^2 \\ + \beta (1 - \beta) K \alpha^2 .
\end{multline}

\noindent Also given that $Var(Z) = K \beta (1 - \beta)$, and $E[Z] = K (1 - \beta)$, we have
\begin{multline}
    Var((1- B) Z) \\ = \alpha (1 - \alpha) \beta (1 - \beta) K + \alpha (1 - \alpha)  K^2 (1 - \beta)^2 \\ + \beta (1 - \beta) K (1 - \alpha)^2 .
\end{multline}

\noindent Additionally,
\begin{multline}
    Cov(BY, (1 - B) Z) \\ =E[BY  (1 - B) Z]  - E[BY] E[(1 - B) Z] 
\end{multline}

\noindent Since $B(1 - B)$ is always zero, we have
\begin{multline}
    Cov(BY, (1 - B) Z) = - E[BY] E[(1 - B) Z] \\ = (\alpha K \beta) [(1 - \alpha) K (1 - \beta)] \\ = -K^2 \alpha (1 - \alpha) \beta (1 - \beta) .
\end{multline}

\noindent Therefore, 
\begin{multline}
    Var(\mathbf{S}) = 2 K \alpha (1 - \alpha) \beta (1 - \beta) + K^2 \alpha (1 - \alpha) (1 - 2 \beta + 2 \beta^2) \\ + K \beta (1 - \beta) (1 - 2 \alpha + 2 \alpha^2) - 2 K^2 \alpha (1 - \alpha) (1 - \beta) \\ = K \beta (1 - \beta) + K^2 \alpha (1 - \alpha) (1 - \beta)^2,
\end{multline}

\noindent and the frequency of $S = 1$, denoted as $\lambda_S$, is represented as
\begin{multline}
    Var(\lambda_S) = \frac{Var(\mathbf{S})}{K^2} \\ = \frac{\beta (1 - \beta)}{K} + \alpha (1 - \alpha) (1 - \beta)^2 .
\end{multline}

\noindent Subsequently, we have

\begin{equation}
    Var(\hat{B'}) = \frac{\beta (1 - \beta)}{K  (1 - \beta)^2} + \alpha (1 - \alpha) .
\end{equation}

Given that the variance of the estimator of the frequency of the original bit $B = 1$ as follows:

\begin{equation}
    Var(\hat{B}) =\frac{\sum_{i \in [N]} Var(\hat{B'}_i)}{N (1 - 2 \alpha)^2} .
\end{equation}

\noindent we have

\begin{equation}
    Var(\hat{B}) = \frac{\beta (1 - \beta)}{NK (1 - 2 \beta)^2 (1 - 2 \alpha)^2} + \frac{\alpha (1 - \alpha)}{N(1 - 2 \alpha)^2} . \qed
\end{equation}

\section{Variance of Estimated Frequency from Randomized Responses}
\label{sec:var_estimated_f}

Suppose an object $X$ has a value $x \in [m]$. An $\epsilon$-LDP randomized response against $X$ generates an output $y \in [m]$, which can also be represented by $m$ random variable $\mathbf{Y} = Y_1, Y_2, ..., Y_m$. If $y$ equals the $i$-th value, then $Y_i = 1$ and other $Y$s are zero. Each variables follow a Bernoulli distribution, and its variance of can be calculated as 

\begin{equation}
    Var(Y_x) = \frac{e^{\epsilon}}{e^{\epsilon} + m - 1} (1 - \frac{e^{\epsilon}}{e^{\epsilon} + m - 1}) ,
\end{equation}

\noindent and

\begin{equation}
\begin{gathered}
    Var(Y_v) = \frac{1}{e^{\epsilon} + m - 1} (1 - \frac{1}{e^{\epsilon} + m - 1}) \\
    \text{s.t. } v \ne x .
\end{gathered}
\end{equation}

Define two more symbols $v \in [m], w \in [m]$ where $v \ne x$ and $w \ne x$. The covariance of $\mathbf{Y}$ is formulated as

\begin{multline}
    Cov(Y_x, Y_v) = E[Y_x  Y_v] - E[Y_x] E[Y_v] \\ =0 - \frac{e^{\epsilon}}{e^{\epsilon} + m - 1} \frac{1}{e^{\epsilon} + m - 1} \\ = - \frac{e^{\epsilon}}{(e^{\epsilon} + m - 1)^2} .
\end{multline}

\noindent and

\begin{multline}
    Cov(Y_v, Y_w) = E[Y_v  Y_w] - E[Y_v] E[Y_w] \\ =0 - \frac{1}{e^{\epsilon} + m - 1} \frac{1}{e^{\epsilon} + m - 1} \\ = - \frac{1}{(e^{\epsilon} + m - 1)^2} ,
\end{multline}

\noindent Therefore, the covariance matrix of $\mathbf{Y}$ is represented as
\begin{multline}
    Cov(Y_i, Y_j) = \frac{1}{(e^{\epsilon} + m - 1)^2} \\
    \times 
    \begin{cases}
       e^{\epsilon} (m - 1)  & \text{if } i = j = x  \\
       - e^{\epsilon} & \text{if } i \ne j \text{ and } (i = x \text{ or } j = x)  \\
       e^{\epsilon} + m - 2  & \text{if } i = j \ne x  \\
       -1  & \text{if } i \ne j \text{ and } i \ne x \text{ and } j \ne x  
    \end{cases} .
\end{multline}

\noindent The above covariance is a function of $x$, and we denote it as $Cov(\mathbf{Y} | x)$. Given a group of objects $X_1, X_2, ..., X_n$, the covariance of the histogram of their randomized responses can be calculated as 

\begin{equation}
    Cov(H) = \sum\limits_{i \in [n]} Cov(\mathbf{Y} | X_i) ,
\end{equation}

\noindent where $H$ represents the histogram of the randomized responses.  The original frequency of different values in $[m]$ can  be estimated as $\frac{1}{n} P^{-1} H$.  Given that $Cov(A x) = A Cov(x) A^T$, where $A$ is a matrix and $x$ is an array, we have

\begin{equation}
    Cov(\frac{1}{n} P^{-1} H ) = \frac{1}{n^2} P^{-1} Cov(H) P^{-1} .
\end{equation}

\end{document}